\documentclass[12pt]{article}
\newcommand{\blind}{0}
\usepackage{easyReview}
\usepackage{amsmath,amsthm}
\usepackage{amsfonts}
\usepackage{amssymb}
\usepackage{bbm}
\usepackage{mathtools}
\usepackage{mathrsfs}

\usepackage{url}

\usepackage[version=4]{mhchem}
\usepackage{stmaryrd}

\usepackage{subcaption} 
\usepackage{graphicx}
\usepackage{pgfplots}
\usepackage[all]{nowidow}
\usepackage[utf8]{inputenc}
\usepackage{tikz}
\usepackage{multicol}
\usepackage{algpseudocode,algorithm,algorithmicx}
\usepackage{scalerel}
\usepackage{natbib}
\usepackage[normalem]{ulem}

\usepackage[english]{babel}

\usepackage{mathdots}
\usepackage{yhmath}
\usepackage{cancel}
\usepackage{color}
\usepackage{array}
\usepackage{multirow}
\usepackage{gensymb}
\usepackage{tabularx}
\usepackage{booktabs}
\usetikzlibrary{fadings}
\usetikzlibrary{patterns}
\usetikzlibrary{shadows.blur}

\usepackage{enumerate}
\usepackage{algorithm}
\usepackage{algpseudocode}
\usepackage{paralist}

\newtheorem{theorem}{Theorem}
\newtheorem{lemma}{Lemma}

\newtheorem{assumption}{Assumption}

\newtheorem{remark}{Remark}
\newtheorem{definition}{Definition}

\newcommand{\bG}{\mathbb{G}}

\newcommand{\cG}{\mathcal{G}}

\newcommand{\cK}{\mathcal{K}}
\newcommand{\cJ}{\mathcal{J}}
\newcommand{\cN}{\mathcal{N}}

\newcommand{\cH}{\mathbb{H}}
\newcommand{\wt}{\widetilde}
\newcommand{\wh}{\widehat}
\newcommand{\R}{\mathbb{R}}
\newcommand{\E}{\mathbb{E}}

\DeclareMathOperator*{\var}{Var}

\newcommand{\li}{\langle}
\newcommand{\ri}{\rangle}

\DeclareMathOperator*{\argmin}{arg\,min}
\newcommand{\indep}{\mathrel{\perp\!\!\!\perp}}
\newcommand{\notindep}{\mathrel{\not\!\perp\!\!\!\perp}}

\usepackage[nodisplayskipstretch]{setspace}

\begin{document}

	\def\spacingset#1{\renewcommand{\baselinestretch}%
		{#1}\small\normalsize} \spacingset{1}

	
	\if0\blind
	{
		\title{\bf Doubly Robust Conditional Independence Testing with Generative Neural Networks}
		\author{Yi Zhang,\hspace{.2cm}Linjun Huang,\hspace{.2cm}Yun Yang\hspace{.2cm}and Xiaofeng Shao
			\thanks{Yi Zhang and Linjun Huang are Ph.D. students, Yun Yang is Associate Professor, Xiaofeng Shao is Professor at Department of Statistics, University of Illinois at Urbana-Champaign. Emails: {\tt yiz19@illinois.edu}, {\tt linjunh2@illinois.edu}, {\tt yy84@illinois.edu} and {\tt xshao@illinois.edu}. Yang's research is partially supported by NSF grant DMS-2210717. Shao's research is partially supported by NSF grants DMS-2210002  and DMS-2412833 and research board grants at UIUC.}\\
		}
		\maketitle
	} \fi
	
	\if1\blind
	{
		\bigskip
		\bigskip
		\bigskip
		\begin{center}
			{\LARGE\bf Title}
		\end{center}
		\medskip
	} \fi
	
	\bigskip
	\begin{abstract}

This article addresses the problem of testing the conditional independence of two generic
random vectors $X$ and $Y$ given a third random vector $Z$, which plays an important role in statistical and machine learning applications.  We propose a new non-parametric testing procedure that avoids explicitly
estimating any conditional distributions but instead requires sampling from the two marginal
conditional distributions of $X$ given $Z$ and $Y$ given $Z$. We further propose using a generative
neural network (GNN) framework to sample from these approximated marginal conditional
distributions, which tends to mitigate the curse of dimensionality due to its adaptivity to any low-dimensional structures and smoothness underlying
the data. Theoretically, our test statistic is shown to enjoy a doubly robust property against
GNN approximation errors, meaning that the test statistic retains all desirable properties of the oracle test statistic utilizing the true marginal conditional distributions, as long
as the product of the two approximation errors decays to zero faster than the parametric
rate. Asymptotic properties of our statistic and the consistency of a bootstrap procedure are derived under both null and local alternatives. Extensive numerical
experiments and real data analysis illustrate the effectiveness and broad applicability of our
proposed test.

	\end{abstract}
	
	\noindent%
	{\it Keywords:} Conditional Distribution;  Conditional Independence Test; Double Robustness; Generative Models; Kernel Method; Maximum Mean Discrepancy.
	\vfill

	\spacingset{1.65} 
	
	\section{Introduction}
	
	Conditional independence (CI) testing plays an important role in various areas of statistics and machine learning. For example, it can be used to examine causal relations \citep{Judeacausality}, determine the structure of graphical models \citep{koller2009probabilistic} and simplify statistical models by performing variable selection \citep{george2000variable, azadkia2021simple} and dimension reduction \citep{cook2002dimension, li2018sufficient}. In this paper, we focus on testing if two generic random vectors $X\in\R^{d_X}$ and $Y\in\R^{d_Y}$ are conditionally independent given a third random vector $Z\in\R^{d_Z}$, where $d_X$, $d_Y$ and $d_Z$ are positive integers; that is, we test the null hypothesis $H_0: X\indep Y\,|\,Z$ against $H_1: X\notindep Y\,|\,Z$. In general, CI testing is statistically more challenging than independence testing because it requires comparing the joint conditional distribution with the product of two of its marginals simultaneously for each value realized by the conditioning variable $Z$. For example, as shown in \cite{Shah2020hard}, there does not exist a CI test that can control the type-I error for all joint distributions of $(X,Y,Z)$ under $H_0$ while having nontrivial power against any alternative. 
	Typically, to make the conditional independence testing problem feasible, one must restrict the distribution class by introducing identifiable structures to make conditional distributions statistically estimable; see, for example, \cite{neykov2021minimax}. A common identifiable structure employed in classic literature on conditional distribution estimation (e.g., \cite{bashtannyk2001bandwidth, izbicki2016nonparametric, li2022minimax}) involves imposing smoothness conditions on the conditional densities of $X,Y\,|\,Z=z$ as $z$ varies across its support. However, when $Z$ represents a random high-dimensional complex object, such as images or texts, the conditional distributions of $X,Y\,|\,Z=z$ may not possess a density function; additionally, a smoothness condition alone may be insufficient to capture many low-dimensional intrinsic structures underlying the true conditional distributions.

 \subsection{Related literature}
	The literature on non-parametric CI testing is extensive, and we refer to \cite{li2020nonparametric} for a recent survey. One major challenge in CI testing is to construct a sample version of a population-level CI measure, which typically involves conditional expectations in the form of $\E [\,u(X)\,|\,Z]$ and/or $\E [\,v(Y)\,|\,Z]$ for some real or function valued mappings $u$ and $v$. Examples include CI testing based on conditional characteristic functions \citep{su2007consistent, wang2015conditional} where $u(x)=e^{itx}$ for all $t\in\R$ and conditional cumulative distribution functions \citep{su2014testing, cai2022distribution} where $u(x)=\mathbbm{1}(x\leq t)$ for all $t\in\R$ with $\mathbbm{1}(\cdot)$ denoting the indicator function. Directly estimating these conditional expectations can be inefficient and cumbersome, as it often involves kernel smoothing and numerical integration, which can be difficult to implement and  inaccurate in high-dimensional settings. This requirement of explicit estimation of conditional expectations makes existing methods suffer from the so-called curse of dimensionality: their performance deteriorate drastically as the dimension $d_Z$ of $Z$ and/or the dimensions $(d_X,d_Y)$ of $(X,Y)$ becomes larger \citep[Section 1]{zhou2022deep}. To mitigate the curse of dimensionality on $(d_X,d_Y)$, kernel-based CI tests \citep{fukumizu2007kernel, zhang2012kernel, huang2022kernel, scetbon2022asymptotic, pogodin2024practical, pogodin2022efficient} instead compare the (kernel) mean embeddings of distributions into some reproducing kernel Hilbert space (RKHS) and tend to have better empirical performance even when the dimensions of $(X,Y)$ are high \citep{li2020nonparametric}. However, in the conditional setting, the estimation of the mean embedding of conditional distributions given all values of $Z$, for example, via RKHS-valued regression, still suffers from a curse of dimensionality on the dimension of $Z$ due to a slow convergence rate \citep{li2022optimal} in the relevant norm, i.e.~the Hilbert-Schmidt norm. As a result, when $d_Z$ is from moderate to large, massive auxiliary data, or in the case of data-splitting, a large portion of the sample, needs to be dedicated to estimating these quantities \citep{scetbon2022asymptotic,pogodin2024practical}, resulting in a noticeable loss of testing power; see Appendix in the supplement for more details. 
	
	Another line of research to CI testing, when the limiting null distribution of the test statistic is not pivotal, is through conditional permutation. For example, in \cite{fukumizu2007kernel,neykov2021minimax,kim2022local}, the permutations are performed locally on bins or clusters specified according to the similarity in the values of the conditioning variable $Z$. As a result, this local permutation procedure also requires large sample size and tends to be unreliable when $Z$ is a continuous random vector or the dimension of $Z$ is large \citep[Section 1]{huang2022kernel}. Alternatively, \cite{sen2017model} and \cite{li2023nearest} proposed to use one-nearest neighbor sampling to perform local permutation. However, their approach requires sample splitting (two thirds of the sample are used for one-nearest neighbor sampling) and certain strong smoothness assumptions on both the marginal density of $Z$ as well as the conditional density of $Y$ given $Z$, which are difficult to verify in practice.
	
	From a different perspective, the above-mentioned challenges can be explained as stemming from the inability to sample from the conditional distribution of $X$ and/or $Y$ given $Z$, because in practice we only have i.i.d. copies of $(X,Y,Z)$ and the true conditional distributions are unknown.  With access to a (conditional) generative model for sampling from the conditional distributions, one can easily approximate the conditional expectations, which are the building blocks for most CI test procedures, to arbitrary accuracy without sacrificing computational efficiency using the Monte Carlo method. An (asymptotically) valid rejection criterion with controlled size can then be specified, for example, by utilizing conditional randomization techniques \citep{candes2018panning}, without relying on the knowledge of the limiting null distributions. However, the validity of such conditional randomization-based tests requires exact knowledge of the true (conditional) generative models in order to maintain the desired size, and these tests tend to be quite sensitive to additional errors arising from estimating the practically unknown (conditional) generative models, typically requiring a parametric convergence rate despite the nonparametric nature of (conditional) distribution estimation. Recently, generative neural networks (GNN) based approaches, such as generative adversarial network \citep[GAN,][]{ goodfellow2014generative} and generative moment matching network \citep[GMMN,][]{dziugaite2015training, li2015generative}, have received great success in generating high-quality samples, such as images and texts, from complex and possibly high-dimensional distributions by passing random noise through a deep neural network learned from training samples. In addition to its superior empirical performance, there is also a recent surge of theoretical works on understanding the properties of GNNs. For example, \cite{tang2023minimax} derived the minimax rate for implicitly estimating a high-dimensional distribution that resides on a low-dimensional manifold under a generative model framework, where the rate is shown to depend on the intrinsic dimension and smoothness of the manifold as well as the smoothness of the probability density; \cite{chen2020distribution} showed that GANs are consistent estimators of distributions with certain smooth densities or low-dimensional structures.

	The capacity of GNNs and their success in generating complex random objects mimicking the training samples enable one to approximate various quantities, such as probabilities, quantiles, and expectations, related to conditional distributions by applying the Monte Carlo method using properly trained GNN generators. For example, some recent works such as \cite{shi2021double} and \cite{bellot2019conditional} use GAN to approximate and sample from the conditional distributions in constructing their CI testing procedures. The double GANs-based conditional independence test (DGCIT) proposed in \cite{shi2021double} utilizes double GANs to learn two generators to sample from the conditional distributions of $X$ given $Z$ and $Y$ given $Z$ in order to estimate conditional expectations $\E [u(X)|Z]$ and $\E [v(Y)|Z]$ for certain functions $u$ and $v$ specified in their procedure. In comparison, the generative conditional independence test (GCIT) proposed in \cite{bellot2019conditional} uses the aforementioned conditional randomization approach which only requires training one generator to approximate the conditional distribution of $X$ given $Z$. However, in order to control the type-I error, GCIT requires the approximation error of the generator to decay faster than the $n^{-1/2}$ parametric rate with $n$ denoting the sample size, which is not achievable even in a simple linear regression case \citep[Proposition 1]{shi2021double}. In order to reduce the impact due to the slow nonparametric convergence rate of GAN approximation errors on the testing size and power, the two generators in the test statistic used in DGCIT are integrated in a multiplicative manner such that the approximation error of each generator only need to decay faster than $n^{-1/4}$ \citep[Section 3]{shi2021double}. Despite this desirable double robustness property, DGCIT is designed to test the null hypothesis of weak conditional independence \citep{daudin1980partial}, which is not capable of detecting all conditional dependence within the alternative hypothesis $H_1$ of our considered CI testing problem. As a result, DGCIT only has trivial power against certain alternatives in $H_1$; see Section \ref{sec3_3} for further details.

 \subsection{Our contributions}
	In this paper, we propose a GNN based CI test that can address all above-mentioned challenges. In particular, the proposed test exhibits promising empirical performance when the dimension of $Z$ is large, is doubly robust against approximation errors of conditional distributions and remains powerful under all alternatives in $H_1$. In a nutshell, the proposed test is an empirical version of a population-level conditional independence measure based on the maximum mean discrepancy \citep[MMD,][]{gretton2012kernel} between the distribution of $(X,Y,Z)$ under $H_0$ and its true distribution; as a result, it fully characterizes CI under mild assumptions on the employed kernel functions; see Appendix in the supplement for additional literature review comparing our proposed population-level CI measure with other kernel-based ones. In addition, instead of using RKHS-valued regression, we implicitly estimate the mean embeddings of (marginal) conditional distributions via sampling from GNNs trained though sample splitting and cross-fitting. Since the training loss of the GNNs is closely related to the relevant metric in the proposed test statistic, no explicit requirements are needed on the dimensions of $X$, $Y$ and $Z$ to guarantee size control and good power of the testing procedure, provided the trained GNNs are sufficiently accurate; see Section \ref{sec2_1} and Remark \ref{rmk_2_gmmn} of Section \ref{sec:theory}.  Due to the careful incorporation of the GNNs, our test statistic is shown to enjoy a doubly robust property against GNN approximation errors, meaning that the test statistic will retain all desirable properties of the oracle test statistic utilizing the true marginal conditional distributions, as long as the product of the two approximation errors decays to zero faster than the parametric rate. This doubly robust property allows more flexibility and tolerance in estimating the two marginal conditional distributions, making our testing procedure less sensitive to the GNN approximation errors, which typically suffer from a slower non-parametric rate. Specifically, asymptotic properties of our statistic, as well as the consistency of
    a bootstrap procedure, are derived under both null and local alternatives. Finally, extensive numerical experiments and real data analysis illustrate the effectiveness and broad applicability of our proposed test.

 \subsection{Organization and notations}
	The rest of this paper is organized as follows. In Section \ref{sec2_method}, we first motivate and propose our test statistic for CI testing with a bootstrap calibration procedure, and formally introduce the notation of double robustness property in the context of CI testing. In Section \ref{sec:theory}, we develop asymptotic properties of our test statistic under $H_0$ and local alternatives, as well as the consistency results for the proposed wild bootstrap procedure. Section \ref{sec3_simu} evaluates and compares our proposed test along with other representative methods in the literature via finite sample simulations. In Section \ref{sec5_real}, two real data examples are provided using the Cancer Cell Line Encyclopedia dataset in Section \ref{sec5_1} and the MNIST dataset in Section \ref{sec5_2}, respectively. Section \ref{sec6conclu} provides some concluding remarks. All the proofs, as well as other implementation details, are deferred to the supplement.

	The following notations will be used throughout the paper. For any positive integer $d$ and random vectors $(X^1,X^2,\dots,X^d,Z)$ defined on the same probability space, we use $P_{X^1X^2\cdots X^d}$ and $P_{X^1X^2\cdots X^d|Z}$ to denote the joint distribution of $X^1,X^2,\dots,X^d$ and its conditional distribution given $Z$. Let $\E_Z$ or $\E_{Z\sim Q}$ denote taking expectation with respect to random variable $Z$ with distribution $Q$ and $P_m$ denote the Lebesgue measure on $\R^d$. For positive integer $n$, define $[n]=  \{1,2,\dots,n\}$. For probability measure $\mu(\cdot)$ on $\R^{d}$ and $p\geq 1$, let $L_p(\R^d,\mu) = \{f:\R^{d}\to\R: \int|f(x)|^pd\mu(x){<}\infty\}$. For any Hilbert spaces $A$ and $B$, let $A\otimes B$ denote their tensor product. Given random vectors $a,b\in\R^d$, the Gaussian kernel is defined as $\cK(a,b)=\exp(-\|a-b\|_2^2/\sigma)$ and the Laplacian kernel is defined as $\cK(a,b)=\exp(-\|a-b\|_1/\sigma)$, where $\sigma>0$ is the bandwidth parameter, $\|\cdot\|_2$ is the usual Euclidean ($\ell_2$-)norm and $\|\cdot\|_1$ is the $\ell_1$-norm \citep[Table 3.1]{muandet2017kernel}. 

	\section{CI Testing Using Conditional Generators}\label{sec2_method}
	In this section, we first describe a population-level conditional independence measure and present two of its sample versions, one of which benefits from a double robustness property against possible misspecification of the two marginal conditional distributions. We then propose a sample plug-in test statistic based on the superior version, utilizing the generative neural network technique to sample from the marginal conditional distributions along with data splitting and cross-fitting to remove certain data dependence. Following that, we provide a computational framework and a wild bootstrap method to calibrate the size of the proposed testing procedure. Discussions on the connections between our proposed test statistic and existing kernel-based CI tests in the literature are deferred to Appendix in the supplement.
	
	\subsection{Population-level conditional independence measure}\label{sec2_1}
	
	Recall that $X\in\R^{d_X}$, $Y\in\R^{d_Y}$, $Z\in\R^{d_Z}$ are three random vectors defined on the same probability space, and the goal is to test
 \begin{align}\label{eqn:CI_test}
 H_0: X\indep Y\,|\,Z \quad \mbox{versus}\quad H_1: X\notindep Y\,|\,Z.
 \end{align}
 In order to motivate the construction of our population-level conditional independence measure, it is helpful to introduce two new random vectors $\widetilde X$ and $\widetilde Y$ which satisfy all of the properties in the following lemma (see Appendix in the supplement for proof).

 \begin{lemma}\label{lem:augment}
     Let $(X,Y,Z)\in\R^{d_X\times d_Y\times d_Z}$ be three random vectors with joint distribution $P_{XYZ}$. Then, there exist two random vectors $\widetilde X\in\R^{d_X}$ and $\widetilde Y\in\R^{d_Y}$ such that:
     \begin{enumerate}
         \item $(\widetilde X, \widetilde Y)$ are conditionally independent of $(X,Y)$ given $Z$;
         \item $\widetilde X$ and $\widetilde Y$ are conditionally independent given $Z$;
         \item $P_{\widetilde X|Z=z}=P_{X|Z=z}$ and $P_{\widetilde Y|Z=z}=P_{Y|Z=z}$ for a.e.-$P_Z$ $z\in\R^{d_Z}$. 
     \end{enumerate}
 \end{lemma}

\noindent Note that for any random vectors $\widetilde X$ and $\widetilde Y$ satisfying Property 3 in Lemma~\ref{lem:augment}, we have $P_{\widetilde XZ}=P_{XZ}$ and $P_{\widetilde YZ}=P_{YZ}$, and therefore according to Property 2 of the same lemma, testing problem~\eqref{eqn:CI_test} can be equivalently formulated into the following distribution equality testing problem
\begin{align}\label{eqn:CI_test2}
    H_0: P_{XYZ} = P_{\wt X\wt Y Z} \quad \mbox{versus}\quad H_1: P_{XYZ} \neq P_{\wt X\wt Y Z}.
\end{align}
It is then natural to propose a population-level measure to quantify the discrepancy between $P_{XYZ}$ and $P_{\wt X\wt Y Z}$. In order to mitigate the curse of dimensionality, we adopt a kernel-based measure using MMD, which compares the (kernel) mean embeddings of $P_{XYZ}$ and $P_{\widetilde X \widetilde Y Z}$ in an RKHS. We call the resulting measure as
an MMD-based CI measure, or MMDCI in short. More specifically, we let $\mathcal{K}_X:\, \R^{d_X}\times\R^{d_X}$, $\mathcal{K}_Y:\, \R^{d_Y}\times\R^{d_Y}$ and $\mathcal{K}_Z:\, \R^{d_Z}\times\R^{d_Z}$ denote three symmetric positive definite kernel functions which define three RKHSs $\cH_X$, $\cH_Y$ and $\cH_Z$ over the space of $X$, $Y$ and $Z$, respectively. We also let $\cK_0 = \cK_X\otimes\cK_Y\otimes\cK_Z$ and $\cH_0$ denote the corresponding (product) kernel and its induced RKHS over the product space $\R^{d_X}\times \R^{d_Y}\times \R^{d_Z}$. 


We use the notation $\li\,\cdot\,,\,\cdot\,\ri_{\cH}$ and $\|\cdot\|_{\cH}$ to denote the associated inner product and the induced norm of a generic RKHS $\cH$ with kernel $\cK$.
For a generic random variable $W$ taking values in the domain of $\cH$, its \emph{(kernel) mean embedding} into $\cH$ is denoted as $\E[\cK(W,\,\cdot)]$, which is defined as the unique element in $\cH$ such that $\E [f(W)] = \li f,\,\E[\cK(W,\,\cdot)\ri_{\cH}$ for any $f\in\cH$. With these notations, we can now define the aforementioned MMDCI as
	\begin{align}
		\text{MMDCI}(P_{XYZ})\,{:\,=}&\,\Big\|\,\E\Big[\cK_0\big((X,Y,Z),\,\cdot\big)\Big]  - \E\Big[\cK_0\big((\widetilde X,\widetilde Y,Z),\,\cdot\big)\Big]\Big\|_{\cH_0}^2 \label{eq_def1} \\
		=&\, \E\Big[\cK_0\big((X,Y,Z),\,(X',Y',Z')\big)\Big] -  2\E\Big[\cK_0\big((X,Y,Z),\,(\widetilde X',\widetilde Y',Z')\big)\Big] \notag\\
     &\qquad\qquad\qquad\qquad\qquad + \E\Big[\cK_0\big((\widetilde X,\widetilde Y,Z),\,(\widetilde X',\widetilde Y',Z')\big)\Big], \label{eqq22}
	\end{align}
where $(X',Y',\widetilde X',\widetilde Y',Z')$ denotes an independent copy of $(X,Y,\widetilde X,\widetilde Y,Z)$, and the second equality~\eqref{eqq22} follows by expanding the squared RKHS norm, using the definition of the mean embedding and applying the reproducing property of the RKHS $\cH_0$. Note that this MMDCI measure only depends on the joint distribution $P_{XYZ}$, as $P_{\wt X\wt Y Z}$ is implicitly defined through $P_{XYZ}$ using Lemma~\ref{lem:augment}. In particular, we have the following Lemma \ref{lem:MMDCI} (which is proved in Appendix of the supplement) relating $\text{MMDCI}(P_{XYZ})$ to $H_0$ in problem~\eqref{eqn:CI_test} or~\eqref{eqn:CI_test2} if Assumption \ref{assump1} holds.

	\begin{assumption}\label{assump1}
		For each $\widetilde Q\in\{\widetilde X, X\}$, assume either one of the following assumptions holds:
		\begin{enumerate}[(a).]
			\item\label{assump1_a} $\E\Big[\sqrt{\cK_X(\widetilde Q,\widetilde Q)\,\cK_Y(Y,Y)\,\cK_Z(Z,Z)}\Big] {<}\infty$. $\cK_0$ is a characteristic kernel \citep[][Section 2.2]{fukumizu2007kernel}, meaning that for any probability measures $P$ and $Q$ on $\R^{d_X+d_Y+d_Z}$,\newline $\E_{(X,Y,Z)\sim P}\big[\cK_0\big((X,Y,Z),\,\cdot\big)\big]=\E_{(X,Y,Z)\sim Q}\big[\cK_0\big((X,Y,Z),\,\cdot\big)\big]$ implies $P=Q$.
			\item\label{assump1_b} $\cH_X \otimes \cH_Z$ is dense in $L_2(\R^{d_X+d_Z},P_{XZ})$ and $\cH_Y$ is dense in $L_2(\R^{d_Y},P_{Y})$.
			\item \label{assump1_c} $\E\big[\cK_X(\widetilde Q,\,\widetilde Q)\,\cK_Y(Y,\,Y)\big]{<}\infty$,  $\cH_Z$ is dense in $L_2(\R^{d_Z},P_{Z})$ and $(\cK_X,\cK_Y)$ are continuous characteristic kernels.
		\end{enumerate}
	\end{assumption}
	\begin{lemma}\label{lem:MMDCI}
	Let $(X,Y,Z){\in}\R^{d_X\times d_Y\times d_Z}$ be three random vectors with joint distribution $P_{XYZ}$. Under Assumption~\ref{assump1}, $X$ and $Y$ are conditionally independent of $Z$  \emph{if and only if} $\text{MMDCI}(P_{XYZ}) = 0$. 
	\end{lemma}
	\begin{remark}\label{rmk2_assumpb}
		Part (\ref{assump1_a}) of Assumption \ref{assump1} comes from the definition of MMDCI in equation (\ref{eq_def1}). Part (\ref{assump1_b}) of Assumption \ref{assump1} comes from the CI characterization in \cite{daudin1980partial}, and it is implied by $\cK_X\otimes \cK_Z$ and $\cK_Y$ being $L_2$- or $c_0$-universal kernels \citep{sriperumbudur2011universality} and it can be alternatively stated as: $\cH_Y \otimes \cH_Z$ is dense in $L_2(\R^{d_Y+d_Z},P_{YZ})$ and $\cH_X$ is dense in $L_2(\R^{d_X},P_{X})$. According to \cite{sriperumbudur2011universality} and Theorems 4 and 5 in \cite{szabo2018characteristic}, Assumption \ref{assump1} holds if $(\cK_X,\cK_Y,\cK_Z)$ are Gaussian or Laplacian kernels. Part (\ref{assump1_a}) is in general more restrictive than part (\ref{assump1_c}) because it implies $(\cK_X,\cK_Y,\cK_Z)$ are all characteristic kernels. If $(\cK_X,\cK_Y,\cK_Z)$ are bounded, continuous and translation invariant kernels, both part (\ref{assump1_a}) and (\ref{assump1_c}) are equivalent to $(\cK_X,\cK_Y,\cK_Z)$ being characteristic kernels \citep[][Theorem 4]{szabo2018characteristic}.
	\end{remark}

In practice, in order to facilitate the construction of a sample version of $\text{MMDCI}(P_{XYZ})$ defined in~\eqref{eqq22}, it is helpful to eliminate the tilde variables $(\wt X,\wt Y, \wt X', \wt Y')$ by taking conditional expectations given $(Z,Z')$ and utilizing the iterative expectation formula and their mutual conditional independence. Specifically, we let $g_X(z){=}\E\big[\cK_X(X,\cdot)\,\big|\,Z{=}z\big]\in \cH_X$ and $g_Y(z){=}\E\big[\cK_Y(Y,\cdot)\,\big|\,Z{=}z\big]\in\cH_Y$ for any $z\in\R^{d_Z}$ denote the \emph{conditional (kernel) mean embedding} of $X$ and $Y$ given $Z$ into their respective RKHS $\cH_X$ and $\cH_Y$. The simplest way to eliminate the tilde variables is to use the factorization property $\cK_0 = \cK_X\otimes\cK_Y\otimes\cK_Z$ along with the iterative expectation formula to \eqref{eqq22} to get (see Appendix in the supplement for a detailed derivation)
	\begin{align}
		\text{MMDCI}(P_{XYZ})
		=&\, \E\Big\{\Big[\cK_X(X,X') \cdot \cK_Y(Y,Y') -  \big\li g_X(Z'),\, \cK_X(X,\,\cdot)  \big\ri_{\cH_X} \cdot \big\li g_Y(Z'),\, \cK_Y(Y,\,\cdot)\big\ri_{\cH_Y} \notag\\
  &\qquad\qquad - \big\li g_X(Z),\, \cK_X(X',\,\cdot)\big\ri_{\cH_X} \cdot  \big\li g_Y(Z),\, \cK_Y(Y',\,\cdot)\big\ri_{\cH_Y} \notag \\
  &\qquad\qquad + \big\li g_X(Z),\,g_X(Z')\big\ri_{\cH_X}  \cdot \big\li g_Y(Z),\,g_Y(Z') \big\ri_{\cH_Y} \Big] \cdot \cK_Z(Z,Z')\Big\} , \label{eqq33}
	\end{align}
 where we deliberately express the quantity inside $\text{MMDCI}(P_{XYZ})$ in a way that is symmetric with respect to $(X,Y,Z)$ and its independent copy $(X',Y',Z')$.
With the knowledge of the two conditional distributions $P_{X|Z}$ and $P_{Y|Z}$, or their mean embeddings $g_X$ and $g_Y$, a sample version of $\text{MMDCI}(P_{XYZ})$ based on equation~\eqref{eqq33}, due to the symmetry, can then be constructed as a degenerate U-statistic of order $2$. Regarding the two conditional distributions, we propose using a GNN framework to approximate their conditional expectations via Monte Carlo sampling, and then plugging-in these conditional expectation estimates into the degenerate U-statistic to construct our sample version of the MMDCI test statistic;  see Section~\ref{sec:sample_ver} for details.

\subsection{A doubly robust population-level measure characterization}\label{sec2_2ora}
	
In order to motivate our construction of a sample version of the MMDCI test statistic that enjoys a desirable double robustness property (see Definition~\ref{def:double_rob}), it is instructive to provide an alternative and equivalent characterization of $\text{MMDCI}(P_{XYZ})$ as summarized in the following lemma, which is proved in Appendix in the supplement. This alternative characterization can be obtained from equation~\eqref{eqq22} by first conditioning on $(Z,Z')$ and then utilizing the factorization property $\cK_0 = \cK_X\otimes\cK_Y\otimes\cK_Z$ and the reproducing property of the RKHSs.

\begin{lemma}\label{lem:MMDCI_alt}
Let $(X,Y,Z)\in\R^{d_X\times d_Y\times d_Z}$ be three random vectors with joint distribution $P_{XYZ}$ and $(X',Y',Z')$ be its independent copy. Then the $\text{MMDCI}(P_{XYZ})$ defined in equation~\eqref{eqq22} can be equivalently written as
\begin{align}
		&\,\text{MMDCI}(P_{XYZ})=\E \Big[U(X,\,X')\,V(Y,\,Y')\,\cK_Z(Z,\,Z') \Big], \qquad 
 \mbox{with}\label{eq_mmdic2}\\
   &\, U(X,\,X') =\cK_X(X,X')-\big\li g_X(Z),\,\cK_X(X',\cdot)\big\ri_{\cH_X}-\big\li g_X(Z'),\,\cK_X(X,\cdot)\big\ri_{\cH_X}+\big\li g_X(Z),\,g_X(Z')\big\ri_{\cH_X}\nonumber\\
	&\,	V(Y,\,Y') = \cK_Y(Y,Y')-\big\li g_Y(Z),\,\cK_Y(Y',\cdot)\big\ri_{\cH_Y}-\big\li g_Y(Z'),\,\cK_Y(Y,\cdot)\big\ri_{\cH_Y}+\big\li g_Y(Z),\,g_Y(Z')\big\ri_{\cH_Y}. \nonumber
	\end{align}
Recall that $g_X(z){=}\E\big[\cK_X(X,\cdot)\,\big|\,Z{=}z\big]\in \cH_X$ and $g_Y(z){=}\E\big[\cK_Y(Y,\cdot)\,\big|\,Z{=}z\big]\in\cH_Y$ for any $z\in\R^{d_Z}$.
\end{lemma}

This alternative characterization of $\text{MMDCI}(P_{XYZ})$ takes a multiplicative form and eliminates the need of introducing $(\widetilde X,\widetilde Y)$ (and its independent copy). Moreover, its constituting multiplicative factors $U(X,\,X')$ and $V(Y,\,Y')$ are both zero mean random variables and symmetric in $(X,X')$ and $(Y,Y')$, respectively. In particular, under $H_0$, $U(X,\,X')$ and $V(Y,\,Y')$ also become conditionally independent given $(Z,Z')$, implying $\text{MMDCI}(P_{XYZ})$ to be zero under $H_0$. Due to the symmetry, one can also construct a sample version of the MMDCI test statistic based on equation~\eqref{eq_mmdic2} as a degenerate U-statistic of order $2$; see the comments after Equation~\eqref{eqq33} and Section~\ref{sec:sample_ver} for further details.

These desirable properties together would lead to a \emph{double robustness} property of this population-level measure $\text{MMDCI}(P_{XYZ})$ expressed from this alternative characterization~\eqref{eq_mmdic2} against potential misspecification of the two (marginal) conditional distributions $P_{X|Z}$ and $P_{Y|Z}$ in estimating $g_X$ and $g_Y$. More generally, we formally define such a population-level double robustness property for any population-level measure that is defined as functions of $(X,Y,Z,X',Y',Z')$ and conditional expectations in the form of $\E [\,u(X)\,|\,Z]$ and/or $\E [\,v(Y)\,|\,Z]$ for certain mappings $u:\,\R^{d_Z}\to \mathbb U$ and $v:\,\R^{d_Z}\to \mathbb V$ with $(\mathbb U,\mathbb V)$ denoting two generic Hilbert spaces as ranges of $(u,v)$ as follows.
\begin{definition}\label{def:double_rob}
    Consider a generic population-level measure taking the form as $D(P_{XYZ})=\\\mathbb E\big[F\big(X,Y,Z,X',Y',Z', \,\E [\,u(X)\,|\,Z], \,\E [\,u(X)\,|\,Z'], \,\E [\,v(Y)\,|\,Z],\,\E [\,v(Y)\,|\,Z']\big)\big]$ for some known mappings $u:\,\R^{d_Z}\to \mathbb U$, $v:\,\R^{d_Z}\to \mathbb V$, and $F:\,\R^{2d_X+2d_Y+2d_Z} \times \mathbb U^2 \times \mathbb V^2 \to \R$. We say that $D(P_{XYZ})$ is doubly robust against misspecifications of the two conditional expectations $g_u(\cdot)=\E [\,u(X)\,|\,Z=\cdot\,]$ and $g_v(\cdot)=\E [\,v(Y)\,|\,Z=\cdot\,]$ if under $H_0$ of testing problem~\eqref{eqn:CI_test}, $D(P_{XYZ})$ equals to zero as long as at least one of these two conditional expectations is correctly specified; that is,  $\mathbb E\big[F\big(X,Y,Z,X',Y',Z', \,g'_u(Z),\,g'_u(Z'),\,g'_v(Z),\,g'_v(Z')\big)\big]=0$ if either $g'_u=g_u$ or $g'_v=g_v$.
\end{definition}

Double robustness is in general a desirable property in many statistical problems where certain estimators or testing procedures maintain their consistency or validity when either of two involved models is correctly specified, but not necessarily both. This property is particularly useful in causal inference and missing data analysis~\citep[see, e.g.,][]{bang2005doubly,funk2011doubly} which involves a treatment model and an outcome model. In our definition of double robustness for CI testing, the two involved models are the two approximation families for the two conditional distributions $P_{X|Z}$ and $P_{Y|Z}$ (or more precisely, the conditional expectations $g_u$ and $g_v$ for certain transformations $u$ and $v$). A theoretical consequence of this population-level double robustness property from Definition~\ref{def:double_rob} on the asymptotic analysis of a sample version of the test statistic is that, the test statistic will retain all desirable properties of the oracle test statistic, defined as the one using the true marginal conditional distributions, as long as the product of the two statistical errors for estimating  $P_{X|Z}$ and $P_{Y|Z}$ (or the relevant conditional expectations $g_u$ and $g_v$) decays to zero faster than the parametric rate; see Section~\ref{sec:theory} for details.
This doubly robust property provides a safeguard against model misspecification, and allows more flexibility and tolerance in estimating the
two marginal conditional distributions, making our testing procedure less sensitive to
the estimation errors, which typically suffer from a slower non-parametric decay
rate.

In our current context, it is straightforward to verify that $\text{MMDCI}(P_{XYZ})$ in Lemma \ref{lem:MMDCI_alt} takes the form as in Definition~\eqref{def:double_rob}; moreover, it is doubly robust against misspecifications of the two conditional mean embeddings $g_X(z){=}\E\big[\cK_X(X,\cdot)\,\big|\,Z{=}z\big]\in \cH_X:\,= \mathbb U$ and $g_Y(z){=}\E\big[\cK_Y(Y,\cdot)\,\big|\,Z{=}z\big]\in\cH_Y:\,=\mathbb V$ since $U(X,\,X')$ and $V(Y,\,Y')$ are both zero mean random variables that are conditionally independent given $(Z,Z')$ under $H_0$. In comparison, under $H_0$ the $\text{MMDCI}(P_{XYZ})$ in its earlier representation~\eqref{eqq33} can also be written in the form as in Definition~\eqref{def:double_rob}; however, it is not doubly robust against misspecifications of the two mean embeddings. To see this, consider replacing $g_X(z)$  in representation~\eqref{eqq33} with $\wh g_X(z) = \E\big[\cK_X(\wh X,\cdot)\,\big|\,Z{=}z\big]$, where $\wh X$ is any random vector satisfying $\wh X\indep Y\mid Z$ and $P_{\wh X|Z}\neq P_{X|Z}$; then $\text{MMDCI}(P_{XYZ})$ after the replacement becomes $\big\|\E[\cK_0((X,Y,Z),\cdot)]- \E[\cK_0((\wh X,Y,Z),\cdot)] \big\|_{\cH_0}^2 $, which is the squared MMD between $P_{XYZ}$ and $P_{\wh XYZ}$ and is strictly positive when $\cK_0$ is a characteristic kernel. See also Section~\ref{sec:sample_ver} for discussions on the consequences in the sample testing statistic accuracy, and Section~\ref{sec:theory} for theoretical implications. The major reason for the same population-level measure having or lacking the double robustness property under different representations is attributed to how $(\wt X, \wt Y, \wt X', \wt Y')$ are eliminated by taking conditional expectations (on which terms) given $(Z, Z')$.

Note that our alternative characterization \eqref{eq_mmdic2} generalizes the one derived in \cite{cai2022distribution} in the special case when $X,Y,Z\sim \mbox{Uniform}[0,1]$, $X\indep Z$, $Y\indep Z$ and $(\cK_X,\cK_Y,\cK_Z)$ are the Laplacian kernel. This characterization \eqref{eq_mmdic2} is closely related to the conditional independence measure proposed in \cite{daudin1980partial}, which is also later used in \cite{zhang2012kernel} and \cite{pogodin2022efficient, pogodin2024practical}, to construct kernel-based CI test statistics. See Appendix for more details and comparisons.

\subsection{Sample conditional independence test statistic}\label{sec:sample_ver}
In this subsection, we provide sample versions arising from the two equivalent characterizations~\eqref{eqq33} and~\eqref{eq_mmdic2} for $\text{MMDCI}(P_{XYZ})$, and discuss the impact of, or the lack of, the double robustness property defined in Definition~\ref{def:double_rob} on their respective statistical accuracy.
In practical calculations, it is easier to utilize the following identities, arising from the reproducing property of the RKHS,
\small\begin{align}\label{eqn:useful_id}
    \big\li g_X(Z),\,\cK_X(X',\cdot)\big\ri_{\cH_X} {=} \E_{X} \big[\cK_X(X,X')\big|Z\big] \ \mbox{and}\ \big\li g_X(Z),\,g_X(Z')\big\ri_{\cH_X} {=}\E_{X,X'}\big[\cK_X(X,X')\big|Z,Z'\big].
\end{align}\normalsize
to further simplify terms involving the conditional mean embeddings $g_X$ and $g_Y$ in these characterizations. More concretely, we first consider the case when $P_{X|Z}$ and $P_{Y|Z}$ (or their corresponding conditional mean embeddings) are known, and construct their sample versions as degenerate U-statistics of order $2$. We call these test statistics as \emph{oracle test statistics} since later we will consider the general and practical case where $P_{X|Z}$ and $P_{Y|Z}$ are unknown and estimated from the data. Let  $\{(X_i,Y_i,Z_i)\}_{i=1}^n$ denote i.i.d.~copies of $(X,Y,Z)$ from $P_{XYZ}$.

\medskip
\noindent {\bf Oracle test statistic based on characterization~\eqref{eqq33}.}
It is easy to see that characterization~\eqref{eqq33} of $\text{MMDCI}(P_{XYZ})$ 
leads to the following oracle test statistic,
\small\begin{align}
		T^\ast_0 =& \,\frac{1}{n(n{-}1)}\sum_{k\neq \ell}\cK_Z(Z_k,Z_\ell)\cdot \Big\{  \cK_X(X_k,X_\ell)\cdot \cK_Y(Y_k,Y_\ell) -\E_{X} \big[K_X(X,X_\ell)\,\big|\,Z=Z_k\big]  \nonumber\\
  &\qquad\cdot \E_{Y} \big[K_Y(Y,Y_\ell)\,\big|\,Z=Z_k\big]  - \E_{X} \big[K_X(X,X_k)\,\big|\,Z=Z_\ell\big] \cdot \E_{Y} \big[K_Y(Y,Y_k)\,\big|\,Z=Z_\ell\big]  \nonumber\\
  & + \E_{X,X'}\big[K_X(X,X')\,\big|\,Z=Z_k,Z'=Z_\ell\big] \cdot \E_{Y,Y'}\big[K_Y(Y,Y')\,\big|\,Z=Z_k,Z'=Z_\ell\big]\Big\}.\label{eqq44}
	\end{align}\normalsize

\medskip
\noindent {\bf Oracle test statistic based on characterization~\eqref{eq_mmdic2}.}
Basing on identity~\eqref{eqn:useful_id}, we can define the following statistic from Characterization~\eqref{eq_mmdic2},
\small\begin{align}
		& \,T^\ast =\frac{1}{n(n{-}1)}\sum_{k\neq \ell}U^\ast(X_k,\,X_\ell)\,V^\ast(Y_k,\,Y_\ell)\,\cK_Z(Z_k,\,Z_\ell),  \qquad \mbox{with}\label{eq_mmdic3}\\
  &\, U^\ast(X_k,\,X_\ell) =\cK_X(X_k,X_\ell) - \E_{X} \big[K_X(X,X_k)\,\big|\,Z=Z_\ell\big] -  \E_{X} \big[K_X(X,X_\ell)\,\big|\,Z=Z_k\big] \nonumber\\
  &\qquad\qquad\qquad\qquad\qquad\qquad\qquad\qquad+ \E_{X,X'}\big[K_X(X,X')\,\big|\,Z=Z_k,Z'=Z_\ell\big],\nonumber\\
	&\,	V^\ast(Y_k,\,Y_\ell) =\cK_Y(Y_k,Y_\ell) - \E_{Y} \big[K_Y(Y,Y_k)\,\big|\,Z=Z_\ell\big] -  \E_{Y} \big[K_Y(Y,Y_\ell)\,\big|\,Z=Z_k\big] \nonumber\\
  &\qquad\qquad\qquad\qquad\qquad\qquad\qquad\qquad+ \E_{Y,Y'}\big[K_Y(Y,Y')\,\big|\,Z=Z_k,Z'=Z_\ell\big].\nonumber
	\end{align}\normalsize

 Under $H_0$ and suitable assumptions, both $nT^\ast_0$ and $nT^\ast$ converge in distribution (after rescaling by the same size) to the typical limiting null distribution of a degenerate U statistic, which is a weighted sum of i.i.d. centered~$\chi_1^2$ random variables. However, they have different limiting distributions; see Appendix for an empirical comparison between their limiting null distributions and powers of tests based on $T_0^\ast$ and $T^\ast$. Specifically, we observe that the test based on $T^\ast$ tends to produce a higher (size-adjusted) power than $T_0^\ast$ under conditional dependence since $T^\ast$ tends to exhibit a smaller asymptotic variance.

\subsubsection{Conditional mean embedding estimation via generative neural networks}\label{sec:GNN}
In practice, since $P_{X|Z}$ and $P_{Y|Z}$ are rarely known, we need to find estimators $\widehat g_X(z)$ and $\widehat g_Y(z)$ for the two conditional mean embeddings $g_X(z)\in\cH_X$ and $g_Y(z)\in\cH_Y$ for each $z\in\R^{d_Z}$. In the literature, \cite{pogodin2024practical} propose to estimate them via RKHS-valued kernel ridge regression (see Appendix for details), which may suffer from slow convergence rate due to the curse of dimensionality on $d_Z$ (see Appendix for an empirical comparison). In this work, in order to mitigate this issue, we adopt a generative neural network (GNN) framework to instead train two conditional generators $\widehat \bG_X:\, \R^{m}\times \R^{d_Z}\to \R^{d_X}$ and $\widehat \bG_Y :\, \R^{m}\times \R^{d_Z}\to \R^{d_Y}$ for approximately sampling from the respective conditional distributions $P_{X|Z}$ and $P_{Y|Z}$. Concretely, if one sample a latent variable $\eta$ from some simple distribution (such as the standard normal) over the latent space $\R^{m}$ for some integer $m\geq 1$, then the conditional distributions of $\widehat X = \widehat \bG_X(\eta,Z)$ and $\widehat Y = \widehat \bG_Y(\eta,Z)$ given $Z$ (denoted as $P_{\widehat X|Z}$ and $P_{\widehat Y|Z}$) are good approximations of $P_{X|Z}$ and $P_{Y|Z}$. In fact, according to the noise-outsourcing lemma (see Theorem 6.10 of \citet{kallenberg2002foundations}, also Lemma 2.1 of \citet{zhou2022deep}), for any integer $m\geq 1$, there exist measurable functions $\bG_X$ and $\bG_Y$ such that for any $\eta\sim N(0,I_m)$ that is independent of $Z$, we have $\bG_X(\eta,Z)\,|\,Z\sim P_{X|Z}$ and $\bG_Y(\eta,Z)\,|\,Z\sim P_{Y|Z}$.
To further estimate $g_X(z)$ for any $z\in\R^{d_Z}$, one can first generate $M$ i.i.d.~samples of $\{\eta_i\}_{i=1}^M$ from $N(0,I_m)$, and then estimate $g_X(z)$ by the sample average $\widehat g_X(z):\,=M^{-1} \sum_{i=1}^{M}\cK_X(\widehat \bG_X(\eta_i,z),\,\cdot\,)\in\cH_X$. Likewise, we can define estimator $\widehat g_Y(z)\in\cH_Y$ for $g_Y(z)$ in a completely analogous manner.

There are many GNN methods available for approximating conditional distributions. We adopt the GMMN framework due to its competitive performance and a close connection (minimizing an MMD) with our proposed test (see Remark \ref{rmk_2_gmmn}).
Concretely, under our notations, the conditional generator $\widehat \bG$ in GMMN for approximating $P_{X|Z}$ ($P_{Y|Z}$ is treated in an analogous way) is obtained by minimizing the sample version of a squared MMD between $P_{XZ}$ and the induced joint distribution $P_{\widehat XZ}$ from the estimated $\widehat X = \widehat \bG_X(\eta,Z)$ based on a generic set of training data $\{(X_i,Z_i)\}_{i=1}^{n_T}$ with training sample size $n_T$ and $Mn_T$ latent variables $\{\eta_i^m:\, i=1,\ldots,n_T,\, m=1,\ldots,M\}$:
\begin{align} 
		\widehat \bG_X & = \argmin_{\bG_X\in \mathcal G_X} \frac{1}{n_T(n_T{-}1)}\sum_{\substack{k\neq \ell\\ k,\ell\in [n_T]}} 	\widehat U(X_k,X_\ell) \cdot \cK_Z(Z_k,Z_\ell),\label{eq_1222}\\
      \mbox{with}\quad 	\widehat U(X_k,X_\ell) & = \cK_X(X_k,X_\ell)-\frac{1}{M}\sum_{m=1}^{M}\cK_X\big(X_k,\,\bG_X(\eta_\ell^{m},Z_\ell)\big)\notag\\
      -\frac{1}{M}\sum_{m=1}^{M}\cK_X&\big(X_\ell,\,\bG_X(\eta_k^{m},Z_k)\big)+\frac{1}{M^2}\sum_{m_1,m_2=1}^{M}\cK_X\big(\bG_X(\eta_k^{m_1},Z_k),\,\bG_X(\eta_\ell^{m_2},Z_\ell)\big), \nonumber 
\end{align}
where $\mathcal G_X$ is an approximation family, such as (deep) neural networks, for the conditional generators. If $\cK_X\otimes \cK_Z$ is a bounded characteristic kernel and there exist $\bG_X^\ast\in \mathcal G_X$ such that $(\bG_X^\ast(\eta,Z),Z)\stackrel{d}{=}(X,Z)$, the generalization bound of GMMN \citep[Theorem 1]{dziugaite2015training} guarantees that the MMD with kernel $\cK_X\otimes \cK_Z$ between the distributions of $(\widehat \bG_X(\eta,Z),Z)$ and $(X,Z)$ converges to zero at the rate $n_T^{-r}$; here, constant $r>0$ typically depends on smoothness properties of $\bG_X^\ast$, intrinsic dimensionality of $X$ and/or $Z$, choices of the kernels and some complexity measure of $\mathcal G_X$, and can be as large as $1/2$ (corresponding to a parametric convergence rate). 
Note that in the original GMMN framework in \cite{dziugaite2015training}, the Monte Carlo sample size $M$ is fixed at one, while our numerical results suggest that choosing a larger $M$ (e.g.~in the order of $n_T$) can dramatically improve the empirical performance and stability of the training process. Specific designs of the neural network architectures for $\mathcal G_X$ in our GMMN implementation and choices of other tuning parameter values in the training process are provided in Appendix.

\subsubsection{Comparison of two oracle test statistics}\label{sec:comparison}
In this part, section we compare the two oracle test statistics \eqref{eqq44} and \eqref{eq_mmdic3} by examining the impact of, or the lack of, the double robustness property, as defined in Definition~\ref{def:double_rob}, on their respective statistical accuracy after plugging-in estimated conditional mean embeddings.

To formally study the impact of plugging-in inaccurate conditional mean embeddings, we consider two generic error metrics $D_X(g_X,\,\widehat g_X)$ and $D_Y(g_Y,\,\widehat g_Y)$ to quantify the estimation errors of the estimators $\widehat g_X$ and $\widehat g_Y$ for the conditional mean embeddings. For example, when all employed kernels are bounded, we can choose $D_X(g_X,g_X') =\big\{\E\big[\|g_X(Z)-g_X'(Z)\|_{\cH_X}^2\big]\big\}^{1/2}$ (recall that for each $z\in\R^{d_Z}$, we have $g_X(z)\in \cH_X$) and 
$D_Y(g_Y,g_Y') =\big\{\E\big[\|g_Y(Z)-g_Y'(Z)\|_{\cH_Y}^2\big]\big\}^{1/2}$ in our theoretical analysis in Section~\ref{sec:theory}.
Recall that we used $n_T$ to denote the GNN training sample size.
Without loss of generality, we may assume $D_X(g_X,\,\widehat g_X)=O(n_T^{-\alpha_1})$ and $D_Y(g_Y,\,\widehat g_Y)=O(n_T^{-\alpha_2})$ for some nonnegative constants $(\alpha_1,\alpha_2)$. Here, $(\alpha_1,\alpha_2)$ characterizes the respective estimation error decay exponent and depends on the GNN training algorithm and properties of the truth $g_X$ and $g_Y$.
Despite the fact that GNN training, such as by the aforementioned GMMN method, is capable of approximating complex distributions, the estimation error of the resulting estimators usually decay slower than the parametric $n_T^{-1/2}$ rate due to its nonparametric nature.

Now suppose that we have two folds of mutually independent samples, with $n_T$ training samples for estimating $(g_X,g_Y)$ and $n$ fitting samples for constructing the two oracle test statistic $T_0^\ast$ and $T^\ast$ as defined in \eqref{eqq44} and \eqref{eq_mmdic3}. 
Let $\widehat T_0 =T_0(\widehat g_X,\widehat g_Y)$ and $\widehat T=T(\widehat g_X,\widehat g_Y)$ denote the respective plug-in test statistic from the oracle ones $T_0^\ast$ and $T^\ast$ obtained by replacing $g_X$ and $g_Y$ (or the corresponding conditional expectations given $Z$ arising from identity~\eqref{eqn:useful_id}) therein with estimators $\widehat g_X$ and $\widehat g_Y$. Using this notation, we can also write $T_0^\ast=T_0(g_X,g_Y)$ and $T^\ast=T(g_X,g_Y)$. Note that we have $\E\big[T_0(g_X,g_Y)\big] = \E\big[T(g_X,g_Y)\big]=0$ under the null $H_0$ due to their constructions. 
Therefore, without any additional structures, a typical perturbation analysis gives that under $H_0$,  
\begin{align*}
	\widehat T_0 & =T_0^\ast+ \underbrace{\Big\{T_0(\widehat g_X,g_Y) - T_0( g_X,g_Y) \Big\}}_{=\,I_1^\ast} \ +\underbrace{\Big\{T_0( g_X,\widehat g_Y) - T_0( g_X,g_Y) \Big\}}_{=\,I_2^\ast} \\
	&\qquad\qquad + \underbrace{\Big\{T_0(\widehat g_X,\widehat g_Y) -T_0(\widehat g_X,g_Y) - T_0( g_X,\widehat g_Y)+T_0( g_X,g_Y)  \Big\}}_{=\,I_3^\ast} \\
	& = T_0^\ast\ + \ O_p\big(n_T^{-\alpha_1}n^{-1/2} + n_T^{-2\alpha_1}\big) + \ O_p\big(n_T^{-\alpha_2}n^{-1/2} + n_T^{-2\alpha_2}\big) + \ O_p\big(n_T^{-\alpha_1-\alpha_2}\big).
\end{align*}
Here, since the MMDCI$(P_{XYZ})$ in characterization~\eqref{eqq33} is not doubly robust, expectations $\big(\E [I_1^\ast],\,\E [I_2^\ast],\,\E [I_3^\ast]\big)$ of the three terms $(I_1^\ast, I_2^\ast,I_3^\ast)$ are generally not zero. As a result, as shown in Appendix, $I_1^\ast$ resembles a non-degenerate U-statistic (with nonzero mean) and satisfies $I_1^\ast = O_p\big(\sqrt{\var(I_1^\ast)}+\E |I_1^\ast|\big)=\ O_p\big(n_T^{-\alpha_1}n^{-1/2} + n_T^{-2\alpha_1}\big)$ (similar results hold for $I_2^\ast$). For the ``higher-order interaction term'' $I_3^\ast$, its decay rate satisfies $I_3^\ast=\ O_p\big(n_T^{-\alpha_1-\alpha_2}\big)$ and depends on both estimation errors. This means that in order for the plug-in error $\widehat T_0 - T_0^\ast$ to be asymptotically negligible for approximating $T_0^\ast$, or $\widehat T_0 - T_0^\ast = o_p(T_0^\ast)$ as $(n_T,n)\to\infty$, so that $\widehat T_0$ can retain all desirable properties of the oracle test statistic $T_0^\ast$, we need $n_T^{-\alpha_1}n^{-1/2} + n_T^{-2\alpha_1} +n_T^{-\alpha_2}n^{-1/2} + n_T^{-2\alpha_2} + n_T^{-\alpha_1-\alpha_2} =o(n^{-1})$, or $\min(\alpha_1,\,\alpha_2)>1/2$ if $n_T$ scales as the same order as $n$. In Appendix, we empirically show that the plug-in test statistic $\widehat{T}_0$ based on $T_0^\ast$ indeed has non-diminishing size distortion, even when both estimators $(\widehat{g}_X, \widehat{g}_Y)$ attain a fast $n^{-1/2}$ rate of convergence when $n_T = n$ in a parametric setting. This implies that our previous perturbation analysis is tight and naively plugging in these estimators into a simple sample version of the population-level CI measure can lead to non-diminishing size distortion in testing $H_0$.

Fortunately, our improved oracle statistic $T^\ast$ as defined in \eqref{eq_mmdic3} admits a better perturbation bound, because of its population-level double robustness property as defined in Definition~\ref{def:double_rob}. More specifically, we can follow the same decomposition to obtain
\begin{align*}
	\widehat T & =T^\ast+ \underbrace{\Big\{T(\widehat g_X,g_Y) - T( g_X,g_Y) \Big\}}_{=\,I_1^\dagger} \ +\underbrace{\Big\{T( g_X,\widehat g_Y) - T( g_X,g_Y) \Big\}}_{=\,I_2^\dagger} \\
	&\qquad\qquad + \underbrace{\Big\{T(\widehat g_X,\widehat g_Y) -T(\widehat g_X,g_Y) - T( g_X,\widehat g_Y)+T( g_X,g_Y)  \Big\}}_{=\,I_3^\dagger} \\
	& = T^\ast\ + \ O_p\big(n_T^{-\alpha_1}n^{-1}\big) + \ O_p\big(n_T^{-\alpha_2}n^{-1} \big) + \ O_p\big(n_T^{-\alpha_1-\alpha_2}n^{-1/2}+n_T^{-2(\alpha_1+\alpha_2)}\big),
\end{align*}
where now the decay rates of the two terms $I^\dagger_1$ and $I^\dagger_2$ have an extra factor of $n^{-1/2}$ compared to those in the earlier decomposition of $\widehat T_0$, due to the double robustness property $\E [I^\dagger_1]=\E [I^\dagger_2]=0$. Specifically, $I^\dagger_1$ now resembles a degenerate U-statistic (with zero mean) and satisfies $I_1^\dagger = O_p\big(\sqrt{\var(I_1^\dagger)}\big)=\ O_p\big(n_T^{-\alpha_1}n^{-1} \big)$ (similar results hold for $I_2^\dagger$). Additionally, although term $I^\dagger_3$ does not have a zero mean, it is a higher-order remainder term and also has a much faster decay rate than $I_3^\ast$, that is, $I_3^\dagger = O_p\big(n_T^{-\alpha_1-\alpha_2}n^{-1/2}+n_T^{-2(\alpha_1+\alpha_2)}\big)$. This analysis implies that the plug-in test statistic based on $T^\ast$ will have correct asymptotic size and attain the same asymptotic limit as $T^\ast$ as long as the product of the two estimation errors decays faster than $n^{-1/2}$, or $\alpha_1+\alpha_2>1/2$ if $n_T$ scales as the same order as $n$. This requirement on the conditional distribution estimation accuracy is much weaker than the $\min(\alpha_1,\,\alpha_2)>1/2$ requirement for the previous oracle statistic $T^\ast_0$, and allows both conditional mean embedding estimators $\widehat g_X$ and $\widehat g_Y$ to have nonparametric convergence rates slower than $n^{-1/2}$. For this reason, in the rest of this paper, we will exclusively focus on the improved oracle statistic $T^\ast$ and its plug-in version.

In the context of CI testing, a similar double robustness phenomenon is also observed in other works, albeit against different plug-in objects. For example, \cite{shi2021double} demonstrate this property in their method to test a null hypothesis of weak conditional independence (see Section~\ref{sec3_3} for an empirical illustration),
and \cite{cai2022distribution} propose a testing procedure based on conditional cumulative distribution functions (cdf). Although the testing procedure in \cite{cai2022distribution} is also able to detect all types of conditional dependence as ours, they require both estimation errors of their conditional cdfs estimators based on kernel smoothing to decay faster than $n^{-1/4}$, which is a stronger requirement than our condition of $\alpha_1 + \alpha_2 > 1/2$. Moreover, because of the estimation by kernel smoothing, their method mainly addresses problems with low-dimensional $Z$ and their theory needs to impose strong smoothness assumptions on the conditional densities of $X$ and $Y$ given $Z$. See Section \ref{sec3_shi} for empirical comparison of their method with ours when the dimension $d_Z$ of $Z$ is from moderate to high.

\subsection{Computation and size calibration via wild bootstrap}\label{sec:comp_boot}
The analysis in Section~\ref{sec:comparison} requires the training samples for estimating the conditional mean embeddings to be independent of the testing samples for constructing the sample test statistic. This motivates us to adopt a sample splitting and cross-fitting framework to remove dependence on conditional generator estimation and improve testing size accuracy. In this framework, we divide the entire dataset into $J$ folds, using $(J{-}1)$ folds to train conditional generators and the remaining fold to construct a plug-in test statistic, and finally averaging the statistics over all $J$ folds; see Section \ref{sec3_fold} for a empirical study of the impact of fold number $J$ on the performance of the test statistic.

To be specific, let $(\eta,\kappa)$ be two independent random variables from the latent variable distribution $N(0,I_m)$ on $\R^m$ that are also independent of $(X,Y,Z)$. Recall that we used $\bG_X^\ast$ and $\bG_Y^\ast$ to denote true conditional generators (see Section~\ref{sec:GNN}) that can generate samples from the two conditional distributions $P_{X|Z}$ and $P_{Y|Z}$. It is then easy to verify that $\wt X= \bG_X^\ast(\eta,Z)$ and $\wt Y = \bG_Y^\ast(\kappa,Z)$ serve as a valid pair of random vectors satisfying all properties in Lemma~\ref{lem:augment}.
Let $\mathcal D:\,=\big\{(X_i,Y_i,Z_i)\big\}_{i=1}^n$ denote the dataset with sample size $n$, and $\big\{(\eta_i^m,\kappa_i^m):\, i\in[n],\,m\in[M]\big\}$ be $nM$ i.i.d.~copies of $(\eta,\kappa)$ such that $\big\{(X_i,Y_i,Z_i)\big\}_{i=1}^n$, $\big\{(\eta_i^1,\kappa_i^1)\big\}_{i=1}^n,\dots,\,\big\{(\eta_i^M,\kappa_i^M)\big\}_{i=1}^n$ are mutually independent. We divide $\{1,2,\dots,n\}$ into $J$ equal folds $\mathcal{J}^{(1)},\dots,\cJ^{(J)}$ and let $\cJ^{(-j)}=\{1,2,\dots,n\}\setminus\cJ^{(j)}$. For each $j\in[J]$, we use $\widehat \bG_{X}^{(j)}$ and $\widehat \bG_{Y}^{(j)}$ to denote the conditional generators trained through~\eqref{eq_1222} using data from $\mathcal D_{-j}:\,=\big\{(X_i,Y_i,Z_i)\big\}_{i\in \cJ^{(-j)}}$. For each $j\in[J]$, $i\in\cJ^{(j)}$ and $m\in[M]$, let $\widehat X_i^{(m)} = \widehat \bG_{X}^{(j)}(\eta_i^m,Z_i) $ and $\widehat Y_i^{(m)} = \widehat \bG_{Y}^{(j)}(\kappa_i^m,Z_i) $ if $i\in\cJ^{(j)}$
. We can then define our final test statistic (based on the doubly robust oracle test statistic) as
	\begin{align} \label{eq: test_statistic}
		\widehat T_J = \frac{1}{J}\sum_{j=1}^J\Bigg\{\frac{1}{\frac{n}{J}(\frac{n}{J}-1)}\sum_{\substack{k\neq \ell\\ k,\ell\in\cJ^{(j)}}} 	\widehat U(X_k,X_\ell) \,\widehat V(Y_k,Y_\ell) \,\cK_Z(Z_k,Z_\ell) \Bigg\},
	\end{align}
	where
	\small\begin{align}
		\widehat U(X_k,X_\ell) = &\cK_X(X_k,X_\ell)-\frac{1}{M}\sum_{m=1}^{M}\cK_X(X_k,\widehat X^{(m)}_\ell)-\frac{1}{M}\sum_{m=1}^{M}\cK_X(X_\ell,\widehat X^{(m)}_k)+\frac{1}{M^2}\sum_{m_1,m_2=1}^{M}\cK_X(\widehat X^{(m_1)}_k,\widehat X^{(m_2)}_\ell),\nonumber \\
		\widehat V(Y_k,Y_\ell) = &\cK_Y(Y_k,Y_\ell)-\frac{1}{M}\sum_{m=1}^{M}\cK_Y(Y_k,\widehat Y^{(m)}_\ell)-\frac{1}{M}\sum_{m=1}^{M}\cK_Y(Y_\ell,\widehat Y^{(m)}_k)+\frac{1}{M^2}\sum_{m_1,m_2=1}^{M}\cK_Y(\widehat Y^{(m_1)}_k,\widehat Y^{(m_2)}_\ell). \nonumber
	\end{align}\normalsize
In Section \ref{sec:theory}, we will show that under $H_0$ and suitable assumptions, $n\widehat T_J$ converges in distribution to the sum of $J$ independent variables, each of which follows a typical limiting distribution of a degenerate U statistic. Moreover, thanks to the double robustness property, the test based on $\widehat T_J$ will have (asymptotically) accurate size if the conditional generators have error decay exponents $\alpha_1$ and $\alpha_2$ such that $\alpha_1 + \alpha_2>1/2$ for some specific distance measures $D_X$ and $D_Y$. This allows for nonparametric rates for conditional mean embedding estimation.

\medskip
\noindent {\bf A wild bootstrap procedure for test calibration.} The limiting null distribution of $n\widehat T_J$ is not pivotal and involves unknown quantities related to eigenvalues of certain operators depending on the kernels and the unknown joint distribution $P_{XYZ}$; see Theorem~\ref{th_null} for details. These unknown quantities make the limiting null distribution (and its quantiles) difficult to approximate in practice. Therefore, we instead adopt a wild bootstrap procedure (see, e.g., Section 2.4 of \cite{zhang2018conditional}) to approximate the null distribution of $n\widehat{T}_J$. More concretely, we use $B$ to denote the bootstrap sample size, and
for each $b=1,2,\dots,B$, we generate $n$ i.i.d.~random multipliers $\{e_{bi}\}_{i=1}^n$ from the standard normal distribution $N(0,1)$. A bootstrap version of $\widehat T_J$ is then defined by injecting the random multipliers as 
	\begin{align}
		\widehat T_J^b = \frac{1}{J}\sum_{j=1}^J\Bigg\{\frac{1}{\frac{n}{J}(\frac{n}{J}-1)}\sum_{\substack{k\neq \ell\\ k,\ell\in\cJ^{(j)}}} 	\widehat U(X_k,X_\ell)\, \widehat V(Y_k,Y_\ell)\, \cK_Z(Z_k,Z_\ell)\, e_{bk}\,e_{b\ell}\Bigg\}, \label{eqbt}
	\end{align}
 where $\widehat U(X_k,X_\ell)$ and $\widehat V(Y_k,Y_\ell)$ are defined as earlier.
We then reject $H_0$ at level $\gamma\in(0,1)$ if the empirical bootstrap reject proportion given the data $\mathcal D$ falls below $\gamma$, or $B^{-1}\sum_{b=1}^{B}  \mathbbm{1}_{\{\widehat T_J^b >\widehat T_J\}} < \gamma$.

\section{Theoretical Properties}\label{sec:theory}
In this section, we provide asymptotic analysis of the proposed plug-in test statistic $\widehat T_J$ based on the improved oracle statistic $T^\ast$ using sample splitting and cross-fitting, as defined in Section~\ref{sec:comp_boot}. Our asymptotic analysis includes deriving the limiting distribution of the test statistic and proving the bootstrap consistency under the null $H_0$ as well as analyzing the power of the corresponding testing procedure under a sequence of local alternatives $\{H_{1n}\}_{n=1}^\infty$ that lie near and/or above the detection boundary.

\subsection{Limiting distribution and bootstrap consistency under $H_0$}
	Before presenting our theoretical results, we need to make the following assumption on the employed kernels and qualities of conditional generators $\big\{\big(\widehat \bG_X^{(j)},\,\widehat\bG_Y^{(j)}\big)\big\}_{j=1}^J$, implicitly through their generated samples $(\widehat X_i^{(m)},\,\widehat Y_i^{(m)})$ for $i\in[n]$ and $m\in[M]$, trained during sampling splitting. Recall that $\wt X= \bG_X^\ast(\eta,Z)$ and $\wt Y = \bG_Y^\ast(\kappa,Z)$ denote a valid pair of random vectors satisfying all properties in Lemma~\ref{lem:augment}, whose marginal conditional distributions given $Z$ are $P_{X|Z}$ and $P_{Y|Z}$ respectively and we denote $ \widetilde X_{i}^{(m)} = \bG_X^\ast(\eta_i^m,Z_i), \widetilde Y_{i}^{(m)} = \bG_Y^\ast(\kappa_i^m,Z_i)$.
	\begin{assumption}\label{assump2}
	Assume $M\to \infty$ as $n\to \infty$. There exist constants $C_0>0$, $\alpha_1,\alpha_2>0$ satisfying $\alpha_1{+}\alpha_2>\frac{1}{2}$, such that for $D_{i} \in\{X_{i}, \widetilde X_{i}^{(1)},\widehat X_{i}^{(1)}\}$, $E_{i} \in\{Y_{i}, \widetilde Y_{i}^{(1)}, \widehat Y_{i}^{(1)}\}$ and $i\in \{i_1,\dots,i_J\}$ where $i_s\in \cJ^{(s)}$ for each $s\in[J]$, we have:
	\begin{enumerate}[(a)]
		\item  $\E\Big[\cK_X(D_i,D_i)\,\cK_Y(E_i,E_i)\,\cK_Z(Z_i,Z_i)\Big]{<}C_0$.
		\item\label{assump2_b} \small $\sqrt{\E\Big[\Big\|\E\big[\cK_X(\cdot,X_{i})\,\big|\,Z_{i}\big]{-}\E\big[\cK_X(\cdot,\widehat X_{i}^{(1)})\,\big|\,Z_{i}\big]\Big\|_{\cH_X}^2\cdot \Big(\sqrt{\cK_Z(Z_{i},Z_{i})}{+}\cK_Y(E_{i},E_{i})\,\cK_Z(Z_{i},Z_{i})\Big)\Big]}{=}O(n^{-\alpha_1})$ and $\sqrt{\E\Big[\Big\|\E\big[\cK_Y(\cdot,Y_{i})\,\big|\,Z_{i}\big]{-}\E\big[\cK_Y(\cdot,\widehat Y_{i}^{(1)})\,\big|\,Z_{i}\big]\Big\|_{\cH_Y}^2\cdot \Big(\sqrt{\cK_Z(Z_{i},Z_{i})}{+}\cK_X(D_{i},D_{i})\,\cK_Z(Z_{i},Z_{i})\Big)\Big]}{=}O(n^{-\alpha_2})$.\normalsize
	\end{enumerate}
\end{assumption}
\noindent We make two remarks regarding this assumption.
	\begin{remark}
		If $\cK_X,\cK_Y,\cK_Z$ are bounded kernels (without loss of generality, assume they are bounded by $1$), then Assumption \ref{assump2} reduces to $\E\big[\big\|\E[\cK_X(\cdot, X_{i})\,|\,Z_{i}]-\E\big[\cK_X(\cdot,\widehat X_{i}^{(1)})\,|\,Z_{i}]\big\|_{\cH_X}^2\big] = O(n^{-2\alpha_1})$ and $\E\big[\big\|\E[\cK_Y(\cdot, Y_{i})\,|\,Z_{i}]-\E\big[\cK_Y(\cdot,\widehat Y_{i}^{(1)})\,|\,Z_{i}]\big\|_{\cH_Y}^2\big] = O(n^{-2\alpha_2})$. Note that
		\begin{align}
			& \,\E\big[\big\|\E[\cK_X(\cdot, X_{i})\,|\,Z_{i}]-\E\big[\cK_X(\cdot,\widehat X_{i}^{(1)})\,|\,Z_{i}]\big\|_{\cH_X}^2\big] =\E\Big[\Big\{ \sup_{f\in\cH_X:\,\|f\|_{\cH_X}\leq 1}\E\big[f(\widehat X_{i}^{(1)})- f( X_{i}) \,\big|\,Z_{i} \big]\Big\}^2 \Big]\nonumber
			\\
			\leq &\,\E\Big[\Big\{ \sup_{f:\R^{d_X}\to\R:\,\|f\|_{\infty}\leq 1}\E\big[f(\widehat X_{i}^{(1)})- f( X_{i}) \,\big|\,Z_{i} \big]\Big\}^2 \Big]
			= 2 \,\E \big[d^2_{\rm TV}(P_{ \widehat X_{i}^{(1)}|Z_{i}},P_{ X_{i}|Z_{i}})\big], \label{eqn:relation_TV}
		\end{align}
		where $\|\cdot\|_{\infty}$ denotes the function supreme norm, and $d_{\rm TV}(\cdot,\cdot)$ denotes the total variation distance. Here, the inequality in the second line is implied by the fact that $\|f\|_{\infty} = \sup_{x\in\R^{d_X}}|f(x)|=\sup_{x\in\R^{d_X}}|\li f,\cK_X(x,\cdot)\ri_{\cH_X}|\leq \|f\|_{\cH_X}\sqrt{\cK_X(x,x)}\leq \|f\|_{\cH_X}$. Therefore, we can also replace the error metric in Assumption \ref{assump2} from an MMD type distance to the total variation distance, i.e.,  $ \E \big[d^2_{\rm TV}(P_{ \widehat X_{i}^{(1)}|Z_{i}},P_{  X_{i}|Z_{i}})\big] =  O(n^{-2\alpha_1})$ and $ \E \big[d^2_{\rm TV}(P_{ \widehat Y_{i}^{(1)}|Z_{i}},P_{  Y_{i}|Z_{i}})\big] =  O(n^{-2\alpha_2})$, which is a common assumption made in existing works for characterizing qualities of conditional generators, see \cite{zhou2022deep} and \cite{shi2021double}. However, the total variation metric may not be a suitable metric for characterizing the closeness between nearly mutually singular distributions, which happens when data are complex objects such as images or texts exhibiting low-dimensional manifold structures~\citep{tang2023minimax}. In addition, in our theory, we only require the Monte Carlo sample size $M$ for approximating the conditional expectations to diverge to infinity as $n\to\infty$, while existing works such as \cite{shi2021double} require $M$ to be at least proportional to $n$.
	\end{remark}
	\begin{remark}\label{rmk_2_gmmn}
		The conditional generator $\widehat \bG_{X}^{(j)}$ trained via the GMMN framework as described in Section~\ref{sec:GNN} can be viewed as the minimizer of an empirical risk associated with the squared MMD between $P_{XZ}$ and $P_{\widehat XZ}$ defined as
		\begin{align*}
			\big\|\E\big[\cK_X(X_{i},\cdot)\,\cK_Z(Z_{i},\cdot)\big] {-} \E\big[\cK_X(\widehat X^{(1)}_{i},\cdot)\cK_Z( Z_{i},\cdot) \big]    \big\|^2_{\cH_{XZ}} = \Big\{\sup_{f\in\cH_X\otimes \cH_Z:\,\|f\|_{\cH_{XZ}}\leq 1}\E\big[f(\widehat X_{i}^{(1)},Z)- f( X_{i},Z) \big]\Big\}^2,
		\end{align*}
  where $\|\cdot\|_{\cH_{XZ}}$ denotes the RKHS norm associated with kernel $\cK_X\otimes \cK_Z$.
  By comparing this representation with the first equality in \eqref{eqn:relation_TV}, we see that the population-level risk of GMMN is closely connected to and slightly weaker than the distance measure we used in our assumption for characterizing the conditional generator estimation error. The essential difference is that the supremum in our distance measure is inside the expectation. In fact, it can be shown that the quantity in the preceding display is always upper bounded by
  \begin{align*}
       \E\Big[ \big\|\E[\cK_X(\cdot,X_{i})\,|\,Z_{i}]-\E[\cK_X(\cdot,\widehat X_{i}^{(1)})\,|\,Z_{i}]\big\|_{\cH_X}^2\cdot\cK_Z( Z_{i},Z_{i})\Big], 
  \end{align*}
  which follows from a generalized Jensen's inequality \citep[][Appendix A]{Junhyung2020}. For translation invariant kernels, the above quantity is exactly the error metric we used in part (\ref{assump2_b}) of Assumption \ref{assump2}. We expect that estimation error of the conditional generator estimator $\widehat \bG_{X}^{(j)}$ based on the GMMN relative to this slightly stronger error metric is also possible to control; however, its analysis is beyond the scope of the current work, and is left for future study. In addition, our empirical results in Appendix show that conditional generator trained using the GMMN framework with a matching MMD improves the quality of the resulting testing procedure when compared to conditional generator trained via other framework (e.g., GANs with Sinkhorn loss) in terms of size control.
	\end{remark}

\noindent The following theorem provides the limiting null distribution of our statistic, whose proof is provided in Appendix.
	\begin{theorem}\label{th_null}
		Suppose Assumption \ref{assump2} holds, then under $H_0$, as $n\to\infty$ and $M\to\infty$, we have
  \begin{align}
      \widehat T_J - T_J = &O_p\big(n^{-1}[M^{-1/2}+n^{-\alpha_1}+n^{-\alpha_2}]+n^{-1/2-(\alpha_1+\alpha_2)}\big)\ \ \ \mbox{and} \ \ \
      n\widehat T_J\stackrel{D}{\to}T^\dagger=  \sum_{j=1}^{J}T_j^\dagger,\nonumber
  \end{align}
		where $T_J$ is defined in the same way as $ \widehat T_J$ with $\wh U(X_j,X_k),\wh V(Y_j,Y_k)$ replaced by $U(X_j,X_k),V(Y_j,Y_k)$, and $\{T_j^\dagger\}_{j\in[J]}$ are i.i.d.~random variables with $T_1^\dagger = \sum_{s=1}^{\infty}\lambda_s(\chi_s^2-1)$. Here, $\chi_s^2$ are i.i.d.~chi-square random variables with one degree of freedom, and $\lambda_s$'s are eigenvalues of the compact self-adjoint operator on $L_2(\R^{d_X+d_Y+d_Z},P_{XYZ})$ induced by the kernel function $h((X_1,Y_1,Z_1),(X_2,Y_2,Z_2)) {=} U(X_1,X_2)V(Y_1,Y_2)\cK_Z(Z_1,Z_2)$; that is, there exists orthonormal basis $\{f_i(X_1,Y_1,Z_1)\}_{i=1}^\infty$ of $L_2(\R^{d_X+d_Y+d_Z},P_{XYZ})$ such that 
		$$\int h((X_1,Y_1,Z_1),(X_2,Y_2,Z_2))f_i(X_1,Y_1,Z_1)dP_{XYZ}(X_1,Y_1,Z_1) = \lambda_if_i(X_2,Y_2,Z_2).$$
	\end{theorem}
\begin{remark}
    The proof of Theorem \ref{th_null} depends on a crucial lemma in Appendix, which shows the asymptotic order of the difference between $\wh T_J$ and the oracle statistic $T_J$. Since $ \widehat T_J - T_J = o_p(n^{-1})$ if $\alpha_1>0,\alpha_2>0,M\to \infty$ and $\alpha_1+\alpha_2>1/2$, under Assumption \ref{assump2} we only need to show $n T_J\stackrel{D}{\to}T^\dagger$. Note that $T^\dagger$ is not pivotal and contains unknown quantities that are hard to estimate, which motivates the wild bootstrap procedure. 
\end{remark}


Let $\{e_i\}_{i=1}^n$ be a sequence of independent standard normal random variables, we define $\widehat M_J^{nM}$ according to Equation (\ref{eqbt}) with $\{e_{bi}\}_{i=1}^n$ replaced by $\{e_i\}_{i=1}^n$. For a statistic $B_{nM}$ that depends on $\{X_i,Y_i,Z_i,\eta_i^m,\kappa_i^m\}_{i=1,m=1}^{n,M}$, as in Definition 2.1 of \cite{zhang2018conditional}, we say $B_{nM}$ converges in distribution in probability to a random variable $B^\ast$ if for any subsequence $B_{n_kM_k}$, there is a further subsequence $B_{n_{k_j}M_{n_{k_j}}}$ such that $B_{n_{k_j}M_{n_{k_j}}}\big|\{X_i,Y_i,Z_i,\eta_i^m,\kappa_i^m\}_{i,m=1}^{\infty}$ converges in distribution to $B^\ast$ for a.e. $\{X_i,Y_i,Z_i,\eta_i^m,\kappa_i^m\}_{i,m=1}^{\infty}$. Let $\stackrel{D^\ast}{\to}$ denote convergence in distribution in probability. We have the following theorem on bootstrap consistency under $H_0$, which is proved in Appendix.
	\begin{theorem}\label{th_boot_null}
		Suppose Assumption \ref{assump2} holds, then we have $n \widehat M_J^{nM}\stackrel{D^\ast}{\to}T^\dagger$ as $n\to\infty$ under $H_0$.
	\end{theorem}
This theorem shows that the bootstrapping test statistic $\widehat M_J^{nM}$ admits the same limiting distribution (given data) after appropriate resscaling as the one in Theorem~\ref{th_boot_null} for $\widehat T_J$. Therefore, the rejection threshold for  $\widehat T_J$ computed as the (conditional) quantile based on $\widehat M_J^{nM}$ will lead to an asymptotically valid size for the CI test.

\subsection{Power analysis under local alternatives}
To facilitate the power analysis and characterize the conditional dependence detection capability of the proposed testing procedure, we introduce a sequence of local alternatives $\{H_{1n}\}_{n=1}^\infty$ based on perturbing the conditional mean embedding of the null $H_0$. Recall that the triple under $H_0$, denoted as $(X^0, Y^0, Z^0)$ to distinguish it from the triple under local alternatives, satisfies $X^0 \indep Y^0 \mid Z^0$. We consider a sequence of alternative hypothesis consisting of triple $(X^A_{n},Y^0,Z^0)$ such that $X^A_{n}\notindep Y_0\mid Z_0$ and certain difference measure between the mean embeddings of $P_{X^A_n|Y^0,Z^0}$ and $P_{X^A_n|Z^0}$ converges to zero at rate $n^{-\alpha}$ for some index $\alpha\geq 0$; for example, $\alpha=0$ corresponds to a fixed alternative where $X^A_{n}=X^A$ does not change with $n$. 


Since $(Y_0,Z_0)$ in the local alternative $H_{1n}$ is the same as the null, we only need to specify the conditional distribution $P_{X^A_{n}|Y^0,Z^0}$ of $X^A_{n}$ given $(Y^0,Z^0)$. Under Assumption~\ref{assump1}, the conditional mean embedding of $X$ given $(Y,Z)$ uniquely determines $P_{X|YZ}$, it is therefore more convenient to specify $H_{1n}$ through the conditional mean embedding $\E\big[K_X(X^A_n,\cdot)\,\big|\, Y^0,Z^0\big]\in \cH_X$ as follows,
\begin{align}\label{eq:local_alt}
    H_{1n}: \, \E\big[\cK_X(X^A_n,\cdot)\,\big|\, Y^0,Z^0\big] = \E\big[\cK_X(X_0,\cdot)\,\big| \,Z^0\big] + n^{-\alpha}\cG(Y^0,Z^0),
\end{align}
where $\cG:\,\R^{d_Y}\times \R^{d_Z}\to\cH_X$ is any fixed $\cH_X$-valued mapping (not changing with $n$). Note that under the null $H_0$, we have $\E\big[K_X(X^0,\cdot)\,\big|\, Y^0,Z^0\big]
= \E\big[K_X(X^0,\cdot)\,\big|\, Z^0\big]$, corresponding to $\cG\equiv 0$ or the limit of $H_{1n}$ as $n\to\infty$ when $\alpha>0$. Moreover, further restricting $\E\big[\cG(Y^0,Z^0)\,\big|\,Z^0\big]=0$ can preserve the same (conditional) marginal distribution of $X^A_n$ given $Z^0$ as that under the null, that is, $P_{X^A_n|Z^0} =P_{X^0|Z^0}$, which, however, is not necessary.
Under $H_{1n}$, we have 
\begin{align}
\mbox{MMDCI}(P_{X^A_nY^0Z^0})
=&n^{-2\alpha}\,\E \Big[\big\li\cG(Y^0,Z^0),\,\cG(\wt Y^0,\wt Z^0)\mid \wt Z^0] \big\ri_{\cH_X} \!\cdot V(Y^0,\wt Y^0)\cdot\cK_Z(Z^0,\wt Z^0) \Big]
=O\big(n^{-2\alpha}\big),\nonumber
\end{align}
where $\wt Y^0$ and $\wt Z^0$ are independent copies of $Y^0$ and $Z^0$ respectively. Note that $\mbox{MMDCI}(P_{X^A_nY^0Z^0})$ is the squared MMD (relative to the kernel $\cK_0=\cK_X\otimes\cK_Y\otimes\cK_Z$) between the joint distributions $P_{X^A_nY^0Z^0}$ and $P_{\widetilde X^A_nY^0Z^0}$, where $\widetilde X^A_n$ satisfies that $\widetilde X^A_n\indep Y^0\mid Z^0 $ and $(\widetilde X^A_n,Z^0)\stackrel{d}{=}( X^A_n,Z^0)$ so that $(\widetilde X^A_n,Y^0,Z^0)$ satisfies the null $H_0$. Therefore, the rate $n^{-\alpha}$ quantifies the degree of deviation in the local alternative $H_{1n}$ from the null $H_0$ as $n\to\infty$.
In a related work, \cite{cai2022distribution} considered specifying a local alternative $P_{X^A_n|Y^0,Z^0} = P_{X^0|Z^0}+n^{-1/2}(P_{X^A|Y^0,Z^0}{-}P_{X^A|Z^0})$ based on directly perturbing the conditional distribution $P_{X^0|Y^0,Z^0}$ of $X^0$ given $(Y^0,Z^0)$, which can be viewed as a special case of our characterization of local alternatives with $\cG(Y^0,Z^0) = \int_{\cH_X} \cK_X(x,\cdot)\,\big[dP_{X^A|Y^0,Z^0}(x){-}dP_{X^A|Z^0}(x)\big] \in\cH_X$ and $\alpha=1/2$. 


For any fixed $n$, let $\{(X_i,Y_i,Z_i)\}_{i=1}^n$ be i.i.d copies of $(X^A_{n},Y^0,Z^0)$. The following theorem gives the asymptotic properties of our statistic under $H_{1n}$, which is proved in Appendix.
	\begin{theorem}\label{th_alter}
		Suppose Assumptions \ref{assump1} and \ref{assump2} hold, then under $H_{1n}$, as $n\to\infty$:
		\begin{enumerate}
			\item If $\alpha=0$, then $\sqrt{n}(\widehat T_J-c_0)\stackrel{D}{\to}\frac{1}{\sqrt{J}}\sum_{j=1}^{J}\cG_j^{(0)}$, where \small$c_0 = \E\Big\{U(X_1,X_2)V(Y_1,Y_2)\cK_Z(Z_1,Z_2)\Big\}>0\,$\normalsize and $\cG_j^{(0)}$ are independent mean zero normal random variables with variance equal to \small
			$$4\var\Big(\E\Big\{U(X_1,X_2)V(Y_1,Y_2)\cK_Z(Z_1,Z_2)\Big|X_2,Y_2,Z_2\Big\}\Big).$$\normalsize
			\item If $0{<}\alpha{<} 1/2$, then $n^{2\alpha}\widehat T_J\stackrel{p}{\to}c$, where 
   $c = \E\Big\{\li\cG(Y_1,Z_1),\cG(Y_2,Z_2)\ri_{\cH_X}V(Y_1,Y_2)\cK_Z(Z_1,Z_2)\Big\}{>}0.$
			\item If  $\alpha = 1/2$,  then $n\widehat T_J\stackrel{D}{\to}c{+}T^\dagger{+}\frac{1}{\sqrt{J}}\sum_{j=1}^{J}\cG_j$, where $\cG_j$ are independent mean zero normal random variables, possibly correlated with $T_j^\dagger$ in Theorem \ref{th_null}, with variance equal to
	\small\begin{align}
	&4\var\Big(\E\Big\{\big[\li\cG(Y,Z),\cK_X(X',\cdot) {-} \E\big[\cK_X(X',\cdot)\,\big|\,Z' \big] \ri_{\cH_X}{+}\li\cG(Y',Z'),\cK_X(X,\cdot) {-} \E\big[\cK_X(X,\cdot)\,\big|\,Z \big]\ri_{\cH_X}\big] \nonumber \\
&\qquad\qquad\qquad\qquad\qquad\qquad\qquad\cdot V(Y,Y')\cK_Z(Z,Z')\Big|X',Y',Z'\Big\}\Big),\nonumber
 \end{align}\normalsize
 where $(X,Y,Z)$ and $(X',Y',Z')$ are i.i.d copies from $P_{X^0Y^0Z^0}$.
			\item If $\alpha > 1/2$, $n\widehat T_J\stackrel{D}{\to}T^\dagger$.
		\end{enumerate}
		
	\end{theorem}
	
The following theorem shows the asymptotic behavior of the bootstrapping statistic $\widehat M_J^{nM}$, as well as the asymptotic power of our proposed testing procedure, under $H_{1n}$, which is proved in Appendix.

	\begin{theorem}\label{th_boot}
		Suppose Assumption \ref{assump2} holds, then we have that as $n\to\infty$ and $M\to\infty$:
		\begin{enumerate}
			\item Under $H_{1n}$ with $\alpha>0$, $n \widehat M_J^{nM}\stackrel{D^\ast}{\to}T^\dagger$.
			\item Under $H_{1n}$ with $\alpha=0$, $n \widehat M_J^{nM}\stackrel{D^\ast}{\to}T_1 = \sum_{j=1}^{J}\widetilde T_j$,
			where $\widetilde T_j$ are i.i.d random variables with $\widetilde T_1 = \sum_{i=1}^{\infty}\gamma_i(\chi_i^2-1)$, $\chi_i^2$ are i.i.d chi-square random variables with one degree of freedom and $\gamma_i$s are eigenvalues of 
			$h((X_1,Y_1,Z_1),(X_2,Y_2,Z_2)) = U(X_1,X_2)V(Y_1,Y_2)\cK_Z(Z_1,Z_2)$.
		\end{enumerate}
  If we further assume Assumptions \ref{assump1} holds, then under $H_{1n}$: $P(n\widehat T_J\geq M^\ast_{nM,\gamma})\to 1$ if $\alpha{<}1/2$, $P(n\widehat T_J\geq M^\ast_{nM,\gamma})\to P(c{+}T^\dagger{+}\frac{1}{\sqrt{J}}\sum_{j=1}^{J}\cG_j\geq T_{0,\gamma})$ if $\alpha{=}1/2$ and $P(n\widehat T_J\geq M^\ast_{nM,\gamma})\to \gamma$ if $\alpha>1/2$. 
	\end{theorem}
	
	
	According to Theorems \ref{th_null} and \ref{th_boot}, our proposed testing procedure asymptotically achieves the correct size under $H_0$ and is consistent in the sense that it can correctly reject $H_0$ under any (local) alternative $H_1$ lying outside a $n^{-1/2}$-neighborhood of $H_0$. This means that our test is shown to have nontrivial power under alternatives that approach $H_0$ at the rate $n^{-1/2}$, while in a related work \cite{pogodin2024practical} based on kernel smoothing, the consistency of their method is only derived under a fixed alternative.

\section{Simulation Study}\label{sec3_simu}
	In this section, we examine the size and power properties of our proposed test statistic and compare with several existing tests. In Section \ref{sec3_fold}, we investigate the influence of the fold number $J$ on the empirical performance of $\widehat T_J$. In Sections \ref{sec3_shi}, we consider a post nonlinear noise model and examine the performance of $\widehat T_2$ (i.e.~take $J=2$) when the dimension $d_Z$ of $Z$ is large. Section \ref{sec3_3} contains the result for a weakly conditional independent alternative where both GCIT proposed in \cite{bellot2019conditional} and DGCIT proposed in \cite{shi2021double} have trivial power. Under the alternative, the size adjusted power for all bootstrap-based methods examined in this section are calculated according to \cite{dominguez2000size}.
	For all the simulations in Section \ref{sec3_simu} and the two applications in Section \ref{sec5_real}, we opt to use the Laplacian kernel. The bandwith parameter for each kernel is selected according to the median heuristic \citep[][Section 8]{gretton2012kernel}. For example, the bandwith parameter for $j$th fold is defined as $\sigma^X_j = \text{median} \big\{ \| X_i {-} X_s \|_1 :\, i,s\in\cJ^{(j)}\big\}$ for $j\in[J]$. We remark that our empirical result of $\widehat T_J$ is not overly sensitive to the number of synthetic data $M$ and the number of bootstrap replicates $B$, so we simply fix $M{=}100$ and $B{=}1000$.

	\subsection{Sensitivity with respect to the fold number $J$}\label{sec3_fold}
	
	We investigate the impact of fold number $J$ on the performance of $\widehat T_J$ with $J\in\{2,4,8\}$. We consider a data generation process (DGP) where $Z_i{=}e_{3i}$, $Y_i{=} Z_i {+} e_{1i}$, $X_i {= }Z_i{+} \delta_i e_{1i} {+} (1{ -} \delta_i) e_{2i}$ with $\{e_{1i}\}_{i=1}^n$, $\{e_{2i}\}_{i=1}^n$ and $\{e_{3i}\}_{i=1}^n$ being independent samples following the standard normal distribution and $\delta_i \stackrel{i.i.d}{\sim} Bernoulli(p)$, which is also independent of $\{e_{si}\}_{i=1}^n$ for ${s\in[3]}$. The experiments are repeated 1000 times with nominal level being fixed at 5\%. 
	
\begin{figure}[H]
		\begin{subfigure}{0.5\textwidth}
			\includegraphics[width=1\textwidth]{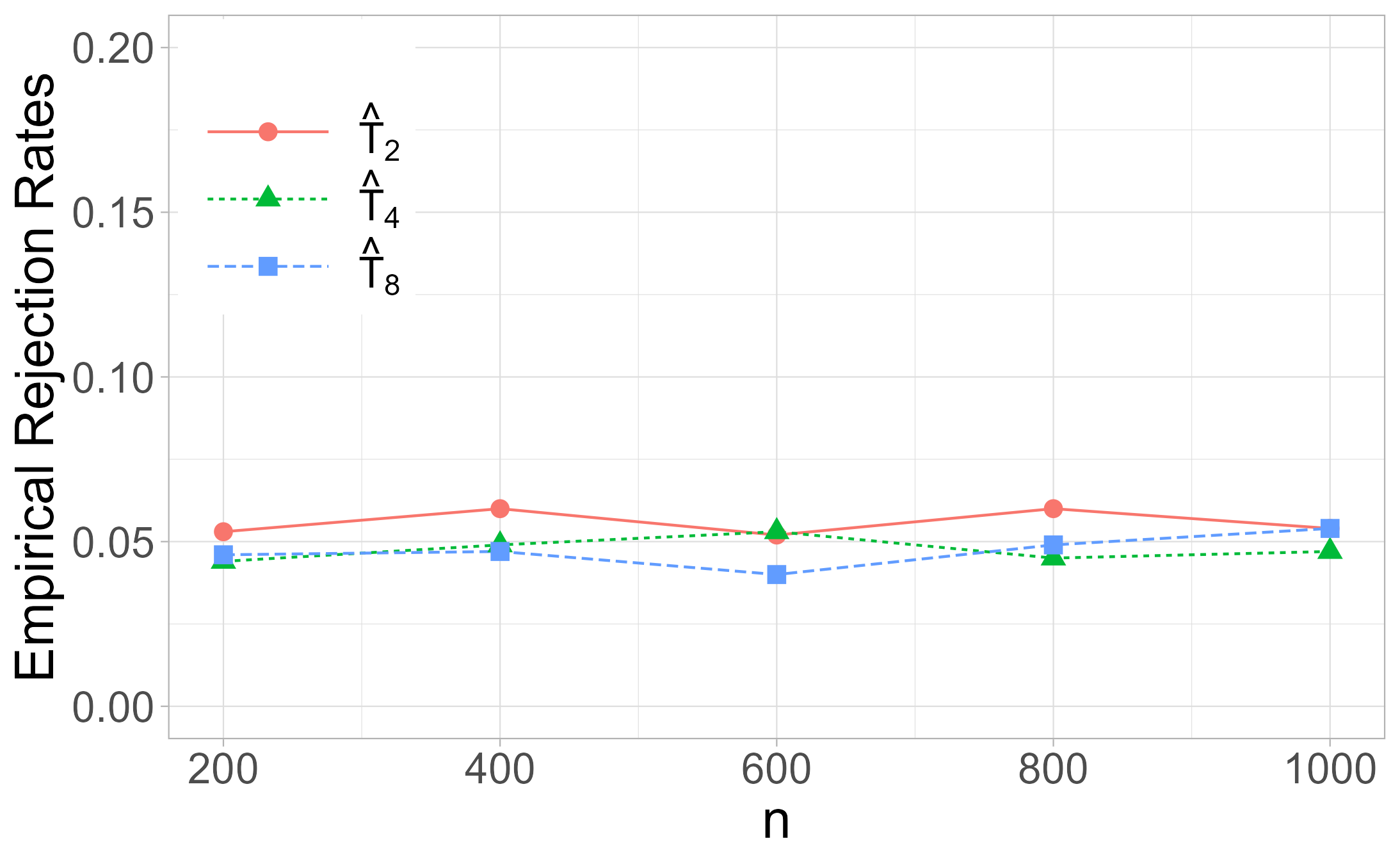}
			\caption{}
			\label{fig_sec3_4_size}
		\end{subfigure}
		\hfill
		\begin{subfigure}{0.5\textwidth}
			\includegraphics[width=1\textwidth]{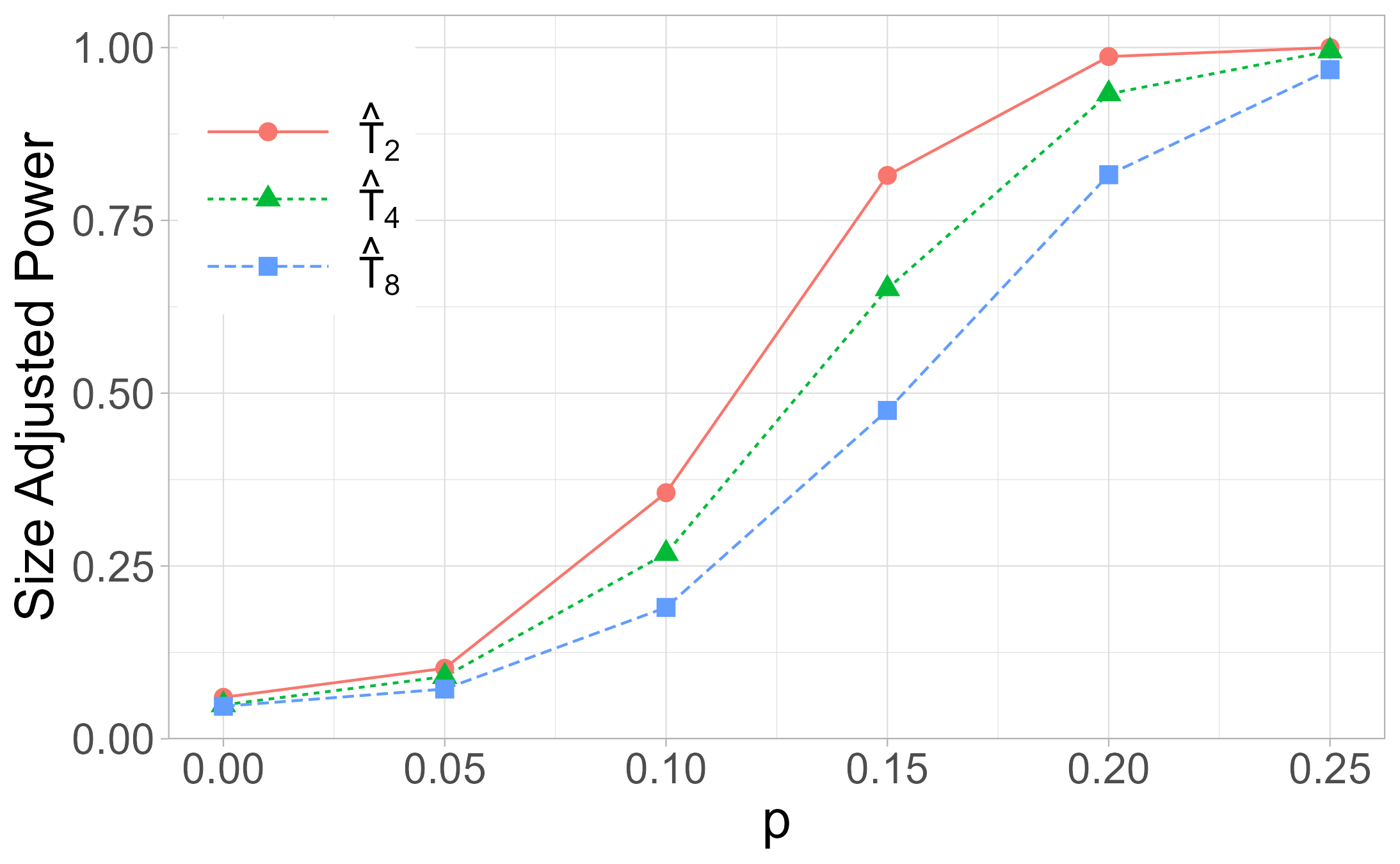}
			\caption{}
			\label{fig_sec3_4_power}
		\end{subfigure}
		\caption{\footnotesize Empirical size (left) and size adjusted power (right) for $\widehat T_J$. }
		\label{fig_sec3_4_1}
	\end{figure}

	For the empirical size, we fix $p=0$ and set $n\in\{200,400,600,800,1000\}$. As shown in Figure \ref{fig_sec3_4_size}, the empirical size for $\widehat T_J$ is close to the nominal level for all $J\in\{2,4,8\}$ for all sample size $n$. For the size adjusted power, we set $n=400$ and plot the power curves against $p\in[0,0.25]$ in Figure \ref{fig_sec3_4_power}. For any fixed $p$, the empirical power for $\widehat T_J$ decreases as $J$ increases. This is expected since only the $J \times\frac{n^2}{J^2}$ entries in the $J$ block diagonal distance matrices of $\{\cK_X(X_j,X_k)\}_{j,k=1}^n$, $\{\cK_Y(Y_j,Y_k)\}_{j,k=1}^n$ and $\{\cK_Z(Z_j,Z_k)\}_{j,k=1}^n$ are used in evaluating $\widehat T_J$ and the number of pairs declines as $J$ increases. Since the empirical size is not sensitive to the choice of $J$ and smaller value of $J$ corresponds to larger power, we recommend setting $J{=}2$ in practice.
	
	 \normalsize

	\subsection{Post nonlinear noise model}\label{sec3_shi}
	
	Consider the following DGP adopted by \cite{bellot2019conditional} and \cite{shi2021double}: $Y_i {= }\sin(a_{fi}^\top Z_i + e_{fi}), X_i {=} \cos(a_{gi}^\top Z_i {+} b Y_i {+} e_{gi})$ and $Z_i\stackrel{i.i.d}{\sim} \cN(0,I_{d_Z})$, where $\{e_{fi}\}_{i=1}^n$, $\{e_{gi}\}_{i=1}^n$ are independent samples from $\cN(0,0.25)$ and the entries of $a_f,a_g\in \R^{d_Z}$ are normalized to the unit $\ell_1$ norm after randomly sampled from Uniform$([0, 1]^{d_Z})$. As comparison, we also include the simulation results for DGCIT, GCIT, the CI test proposed in \cite{cai2022distribution} (denoted as CIT) and an oracle statistic $\widehat T_2$ oracle, which is $\widehat T_2$ with synthetic data generated from the true conditional generators. The experiments are repeated 500 times with sample size $n{=}2000$ and nominal level being fixed at 5\%. 
	
\begin{figure}[H]
		\begin{subfigure}{0.5\textwidth}
			\includegraphics[width=1\textwidth]{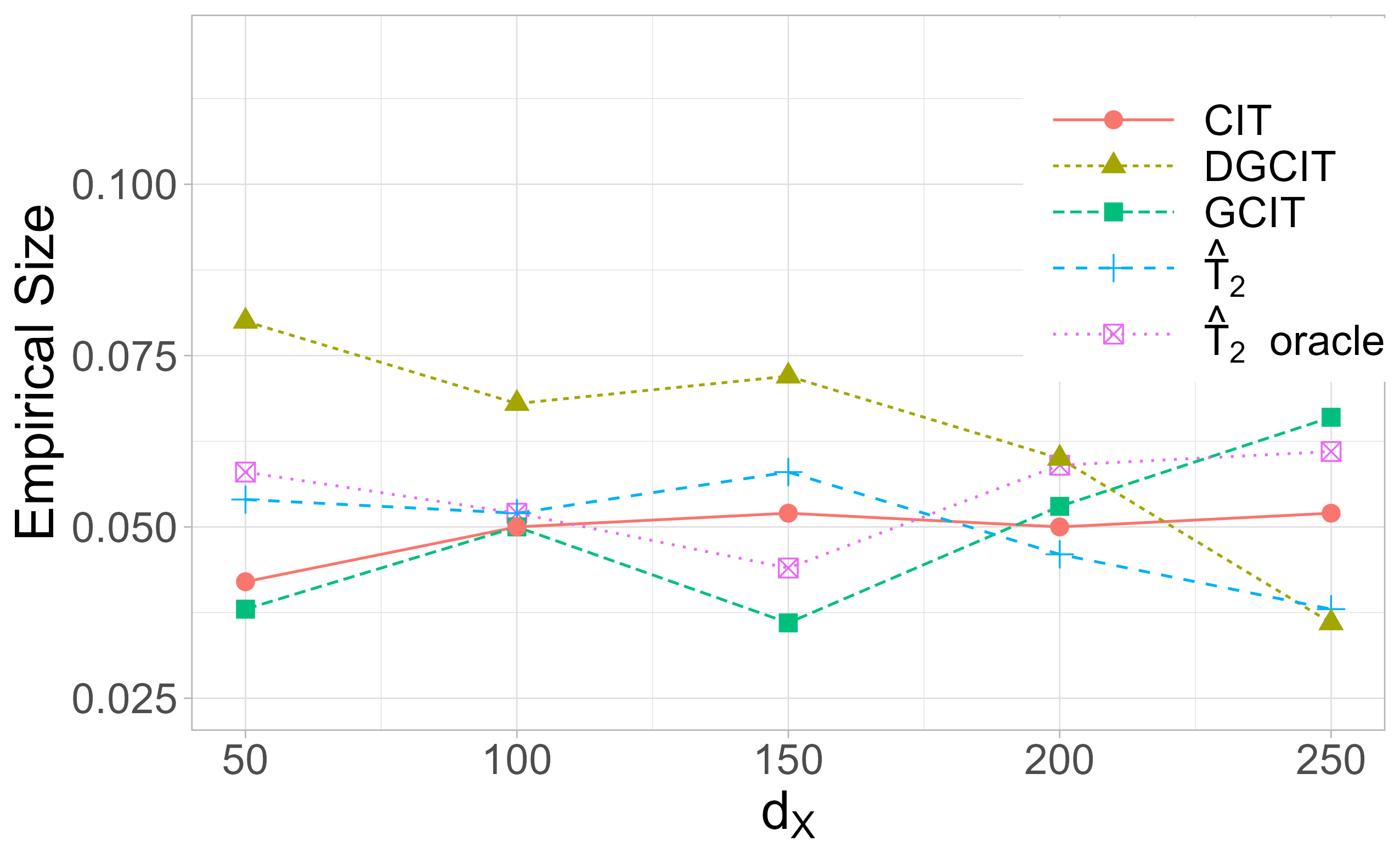}
			\caption{}
			\label{fig_sec3_1_size_5}
		\end{subfigure}
		\hfill
		\begin{subfigure}{0.5\textwidth}
			\includegraphics[width=1\textwidth]{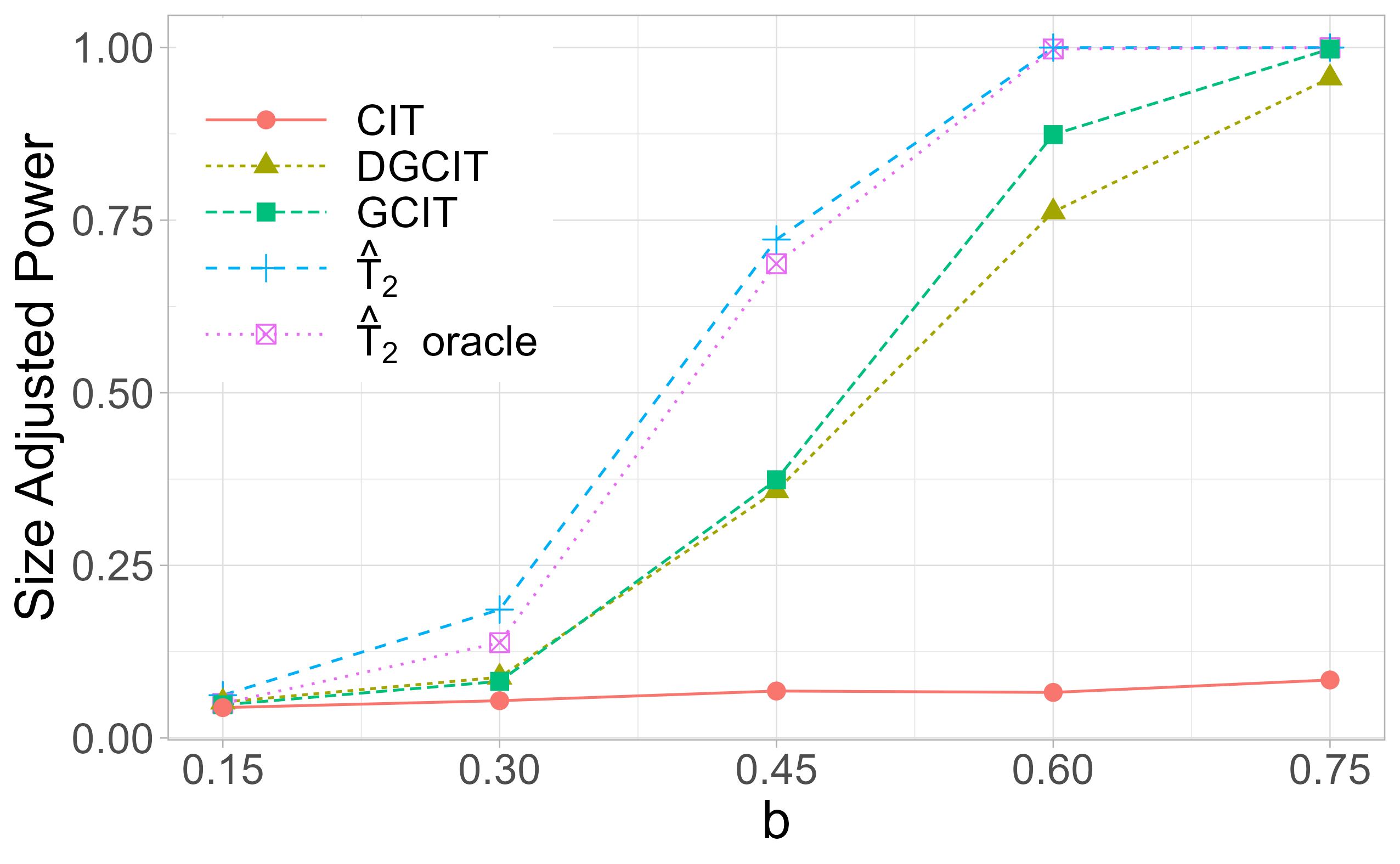}
			\caption{}
			\label{fig_sec3_1_power_5}
		\end{subfigure}
		
		\caption{\footnotesize Empirical size (left) and size adjusted power (right) at nominal level 5\%. }
		\label{fig_sec3_1_1}
	\end{figure}

	For the empirical size, we fix $b=0$ (which corresponds to $H_0$) and set $d_Z\in\{50,100,150,200,250\}$. As shown in Figure \ref{fig_sec3_1_size_5}, $\widehat T_2$ and $\widehat T_2$ oracle have relatively accurate empirical size for $d_Z\leq 200$ and $\widehat T_2$ is mildly undersized when $d_Z{=} 250$. GCIT is undersized when $d_Z\leq 150$ and is oversized when $d_Z=250$. DGCIT goes from oversized to undersized as $d_Z$ increases at 5\% level and it has large size distortion when $d_Z=150$ at level 10\%. For CIT, the empirical sizes are close to the nominal levels for all values of $d_Z$. However, CIT is much more computationally expensive than $\widehat T_2$. As shown in Table \ref{fig_sec3_shi_comp_time}, it takes 18 minutes for CIT to finish one Monte Carlo experiment on an Intel Core i7-11800H CPU when $d_Z{=}50$ and the time increases to 107 minutes when $d_Z=250$. For $\widehat T_2$ the time for one experiment is about 3 minutes for all values of $d_Z$.
	
	For the empirical power, we set $d_Z{=}200$ and plot the power curves against $b\in[0.15,0.75]$ in Figure \ref{fig_sec3_1_power_5}. In general, $\widehat T_2$ has the largest power compared with DGCIT, GCIT and CIT for all values of $b$ and its power curve is close to $\widehat T_2$ oracle. DGCIT outperforms GCIT when $b\leq 0.45$ while the latter has better performance when $b\in\{0.6,0.75\}$. CIT has almost trivial power for all values of $b$. This is expected since it requires iterative estimation of conditional cumulative distribution functions \citep[Section 3.1]{cai2022distribution}, which suffers from the curse of dimensionality and yields poor power performance when the dimension of the conditioning variable is large.

		

	 \normalsize
 
	\begin{table}[H]
            \small
		\vspace{1em}
		\centering
		\begin{tabular}{c|c|c|c|c|c}
			\toprule
			\midrule
			\multirow{2}{*}{}  & \multicolumn{5}{c}{$d_Z$} \\\cline{2-6}
			& $50$ & $100$ & $150$ & $200$ & $250$  \\ \cline{1-6}
			GCIT & $0.50$ & $0.50$ & $0.50$ & $0.50$ & $0.50$  \\\cline{2-6}
			$\widehat{T}_2$ & $2.83$ & $3.00$ & $3.05$ & $3.23$ & $3.3$  \\\cline{2-6}
			DGCIT & $4.85$ & $5.33$ & $5.67$ & $6.23$ & $6.45$  \\\cline{2-6}
			CIT & $18.33$ & $41.27$ & $56.32$ & $80.57$ & $107.35$  \\
			\hline
		\end{tabular} 
		\caption{\footnotesize Computation time for one Monte Carlo experiment (in minutes).  All experiments (except CIT) were conducted on Google Colab using the NVIDIA Tesla T4 GPU. CIT was run on an Intel Core i7-11800H CPU. GCIT used an early stopping criterion.}
		\label{fig_sec3_shi_comp_time}
	\end{table}  \normalsize

	\subsection{Weakly conditional independent alternative}\label{sec3_3}
	
	In this subsection, we show that both DGCIT and GCIT have trivial power against the alternative which is only weakly conditionally independent. Under $H_0$, we generate $\{X_i\}_{i=1}^n$, $\{Y_i\}_{i=1}^n$ and $\{Z_i\}_{i=1}^n$ as mutually independent samples from Bernoulli$(0.5)$. Under the alternative, we consider the same DGP as in Example 5 of \cite{shi2021double} where $Z_i \stackrel{i.i.d}{\sim}Bernoulli(\frac{1}{2})$ and
	\footnotesize\begin{align*}
		&\left( \begin{array}{cc}
			\Pr(X_i=0,Y_i=0|Z_i=0) & \Pr(X_i=0,Y_i=1|Z_i=0) \\
			\Pr(X_i=1,Y_i=0|Z_i=0) & \Pr(X_i=1,Y_i=1|Z_i=0)
		\end{array} \right) = 
		\left( \begin{array}{cc}
			1/6 & 1/3 \\
			1/3 & 1/6
		\end{array} \right), \\
		&\left( \begin{array}{cc}
			\Pr(X_i=0,Y_i=0|Z_i=1) & \Pr(X_i=0,Y_i=1|Z_i=1) \\
			\Pr(X_i=1,Y_i=0|Z_i=1) & \Pr(X_i=1,Y_i=1|Z_i=1)
		\end{array} \right) = 
		\left( \begin{array}{cc}
			1/3 & 1/6 \\
			1/6 & 1/3
		\end{array} \right).
	\end{align*}\normalsize
	
	Under both null and alternative hypothesis, we consider two settings: 1.~the oracle setting where synthetic data are generated from the true conditional generators (with the impact of inaccuracy due to the use of estimated conditional generators being eliminated); 2.~the practical setting where the conditional generators are learned from GNNs. The experiments are repeated 1000 times with sample size $n\in\{200,600,1000\}$ and nominal level being fixed at 5\% and 10\%. As shown in Table \ref{tab_sec3_3_2}, $\widehat T_2$, DGCIT and GCIT have accurate size under $H_0$ for all values of $n$. However, under the alternative in both oracle and practical settings, both DGCIT and GCIT have trivial power while $\widehat T_2$ can successfully detect this alternative.

	\begin{table}[H]
            \small
		\centering
		\setlength{\tabcolsep}{3.5pt}
		\begin{tabular}{c|c|cc|cc|cc|cc|cc|cc}
			\toprule
			\midrule
			& &  \multicolumn{6}{c|}{Null} &  \multicolumn{6}{c}{Alternative} \\\cline{2-14}
			\multirow{2}{*}{} & \multirow{2}{*}{$n$} & \multicolumn{2}{c|}{$\widehat T_2$} &  \multicolumn{2}{c|}{DGCIT} &  \multicolumn{2}{c|}{GCIT}  & \multicolumn{2}{c|}{$\widehat T_2$} &  \multicolumn{2}{c|}{DGCIT} &  \multicolumn{2}{c}{GCIT}     \\
			&	& $10\%$&	$5\%$ &$10\%$&	$5\%$ & $10\%$&	$5\%$ &$10\%$&	$5\%$ &$10\%$&	$5\%$ &$10\%$&	$5\%$    \\  \hline 
			\multirow{3}{*}{Orcale} &$200$ & $10.4$ & $4.3$ & $11.5$ & $6.3$  & $10.3$ & $5.2$ & $96.8$ & $91.6$ & $13.5$ & $7.2$ & $8.3$ & $4$ \ \\ \cline{2-14}
			&$600$ & $10.7$ & $5.5$  & $11.2$ & $6.3$ & $8.2$ & $4.3$ & $100$ & $100$ & $8.5$ & $5$ & $10.1$ & $4.9$ \ \\ \cline{2-14}
			&$1000$ & $9.9$ & $3.9$ & $9.9$ & $4.3$ & $8.2$ & $4.0$ & $100$ & $100$ & $9.8$ & $4.3$  & $10.1$ & $5$  \ \\ \cline{1-14}
			\multirow{3}{*}{Practical} & $200$ & $10.6$ & $4.8$ & $11.2$ & $6.6$ & $12.8$ & $6.2$  & $95.2$ & $89$ & $12.2$ & $7.2$ & $14$ & $6.6$ \ \\ \cline{2-14}
			& $600$ & $9$ & $4.2$  & $11.8$ & $6.6$ & $10.4$ & $4.4$ & $100$ & $100$ & $12.8$ & $7.2$ & $11.6$ & $5.6$ \ \\ \cline{2-14}
			& $1000$ & $11$ & $5.6$ & $12$ & $7$ & $11.2$ & $6.6$ & $100$ & $100$ & $12$ & $7.2$  & $9.4$ & $4.2$   \ \\ \cline{1-14}
		\end{tabular}
		\caption{\footnotesize Empirical rejection rate under null and weakly conditional independent alternative.}\label{tab_sec3_3_2}
	\end{table}  \normalsize

\section{Real Data Applications}\label{sec5_real}
In this section, we apply our method to two real-world datasets: the Cancer Cell Line Encyclopedia (CCLE) dataset \citep{barretina2012cancer} and the MNIST dataset, to assess its empirical performance in specific contexts.
	
	\subsection{CCLE dataset }\label{sec5_1}
	
	The CCLE dataset contains 474 human cancer cell line entries $\{X_i\}_{i=1}^{474}$ and each cell line consists of 1683 genetic mutations taking values zero or one. Here, we follow the same screening procedure as in \cite{shi2021double} and \cite{bellot2019conditional} and consider 466 genetic mutations, so $Z_i{=}(Z_i^1,\dots,Z_i^{466})^\top$. Each cancer cell line $Z_i$ is associated with a response variable $X_i\in \R$, which measures the effect of cancer drug PLX4720. Our goal is to determine if some genetic mutation is related to the response after conditioning on all other mutations. To be specific, assuming $Z_i\stackrel{i.i.d}{\sim}P_Z$, $X_i\stackrel{i.i.d}{\sim}P_X$ and for each $j\in[466]$, we test the null hypothesis $H_0: Z^j\indep X \mid Z^{(-j)}$ against $H_1:  Z^j\notindep X\mid Z^{(-j)}$, where $Z^{(-j)}$ denotes the cancer cell line $Z$ excluding the $j$th mutation.
	
	The p-values for $\widehat T_2$, DGCIT and GCIT for ten different genetic mutations are shown in Table \ref{tab_ccle}, as well as the importance ranking (one means most important) of each mutation obtained from elastic net model (EN, a high-dimensional linear regression with $L_2$ and $L_1$ penalty) and random forest (RF), both of which are commonly used in genetic studies; see \cite{barretina2012cancer}. At 5\% nominal level, both DGCIT and GCIT reject $H_0$ for nine of the ten genetic mutations, with MAP3K5 and FLT3 deemed insignificant by DGCIT and GCIT respectively, while $\widehat T_2$ rejects $H_0$ for all ten mutations. However, as stated in \cite{bellot2019conditional}, PLX4720 is designed as a BARF cancer inhibitor and a proliferation of MAP3K5 is found to be of BRAF type in \cite{prickett2014somatic}. For FLT3, both importance rankings from EN and RF place it as related with $X$ and there is also strong evidence of its importance from genetic research; see \cite{tsai2008discovery} and \cite{larrosa2017flt3}. In conclusion, it is reasonable to believe MAP3K5 and FLT3 are related to $X$ given other mutations and our proposed test $\widehat T_2$ is able to detect both.
	\begin{table}[H]
            \small
		\vspace{1em}
		\centering
		\resizebox{\columnwidth}{!}{\begin{tabular}{ccccccccccc}
				\toprule
				\midrule
				& BRAF.V600E & BRAF.MC & HIP1 & FLT3 & CDC42BPA & THBS3 & DNMT1 & PRKD1 & PIP5K1A & MAP3K5 \\ \hline 
				EN & $1$ & $3$ & $4$ & $5$ & $7$ & $8$ & $9$ & $10$ & $19$ & $78$ \\
				RF & $1$ & $2$ & $3$ & $14$ & $8$ & $34$ & $28$ & $18$ & $7$ & $9$ \\ \hline 
				GCIT & ${<}0.001$ & ${<}0.001$ & $0.008$ & $0.521$ & $0.050$ & $0.013$ & $0.020$ & $0.002$ & $0.001$ & ${<}0.001$ \\
				DGCIT & $0$ & $0$ & $0$ & $0$ & $0$ & $0$ & $0$ & $0$ & $0$ & $0.794$ \\ 
				$\widehat T_2$ & $0$ & $0$ & $0.032$ & $0$ & $0.001$ & $0.048$ & $0.001$ & $0$ & $0.013$ & $0.013$ \\ \hline
		\end{tabular} }
		\caption{ \footnotesize Importance rankings from EN, RF and the p-values from GCIT, DGCIT and $\widehat T_2$.}\label{tab_ccle}
	\end{table} \normalsize

	\subsection{MNIST dataset}\label{sec5_2}
	
	Dimension reduction is a key building block for machine learning applications in various fields, such as computer vision and language processing. Extracting useful features from complex data helps avoid the curse of dimensionality, reduce computation burden and leads to improved model performance \citep{berahmand2024autoencoders}. As a popular non-linear feature extraction methods, the auto-encoder (AE) uses a neural network (encoder) to transform the data $X$ to a $d_l$-dimensional feature vector $f_{d_l}(X)$, which is then transformed back through another neural network (decoder) for reconstruction, resulting in reconstructed data $X'$. A typical AE is trained by minimizing certain reconstruction error, such as the $\ell_2$-norm, between $X$ and $X'$ (after vectorization). The selection of the feature dimension $d_l$ depends on the data and downstream applications and a large $d_l$ may undermine the performance and interpretability of the model \citep{bonheme2022fondue}.
	
	We now demonstrate that our proposed CI test can be used to determine the best value for $d_l$ by using the MNIST dataset that contains 70000 images of handwritten digits $Y=0,1,\dots,9$. Let $\{(X_i,Y_i)\}_{i=1}^{70000}$ denote the $28\times 28$ pixels of images and their corresponding digits, which are assumed to be i.i.d copies of the random vector $(X,Y)$. For different values of $d_l$, we consider testing the null hypothesis $H_0: X \indep Y \mid  f_{d_l}(X)$ against $H_1: X \notindep Y \mid  f_{d_l}(X)$. The minimum value of $d_l$ such that $H_0$ is not rejected will be chosen as the dimension of the feature vector. We split the MNIST dataset into training (with size 60000) and testing (with size 10000) samples. Half of the images in the training dataset are used to train an AE and the other half, as well as its feature vectors from the trained encoder, is used to learn generators of $P_{Y|f_{d_l}(X)}$ and $P_{X|f_{d_l}(X)}$. The test statistic is evaluated on $40$ equal folds of the testing dataset (note that we do not perform sample splitting on each fold). As a comparison, we also include the testing result when the feature vectors are extracted using average pooling \citep[][Section 2.2]{zafar2022comparison} and PCA, which are linear feature extraction methods. For average pooling and PCA, all the images in the training dataset are used to learn $P_{Y|f_{d_l}(X)}$ and $P_{X|f_{d_l}(X)}$.
	
\begin{figure}[H]
		\centering
		\includegraphics[width=0.60\textwidth]{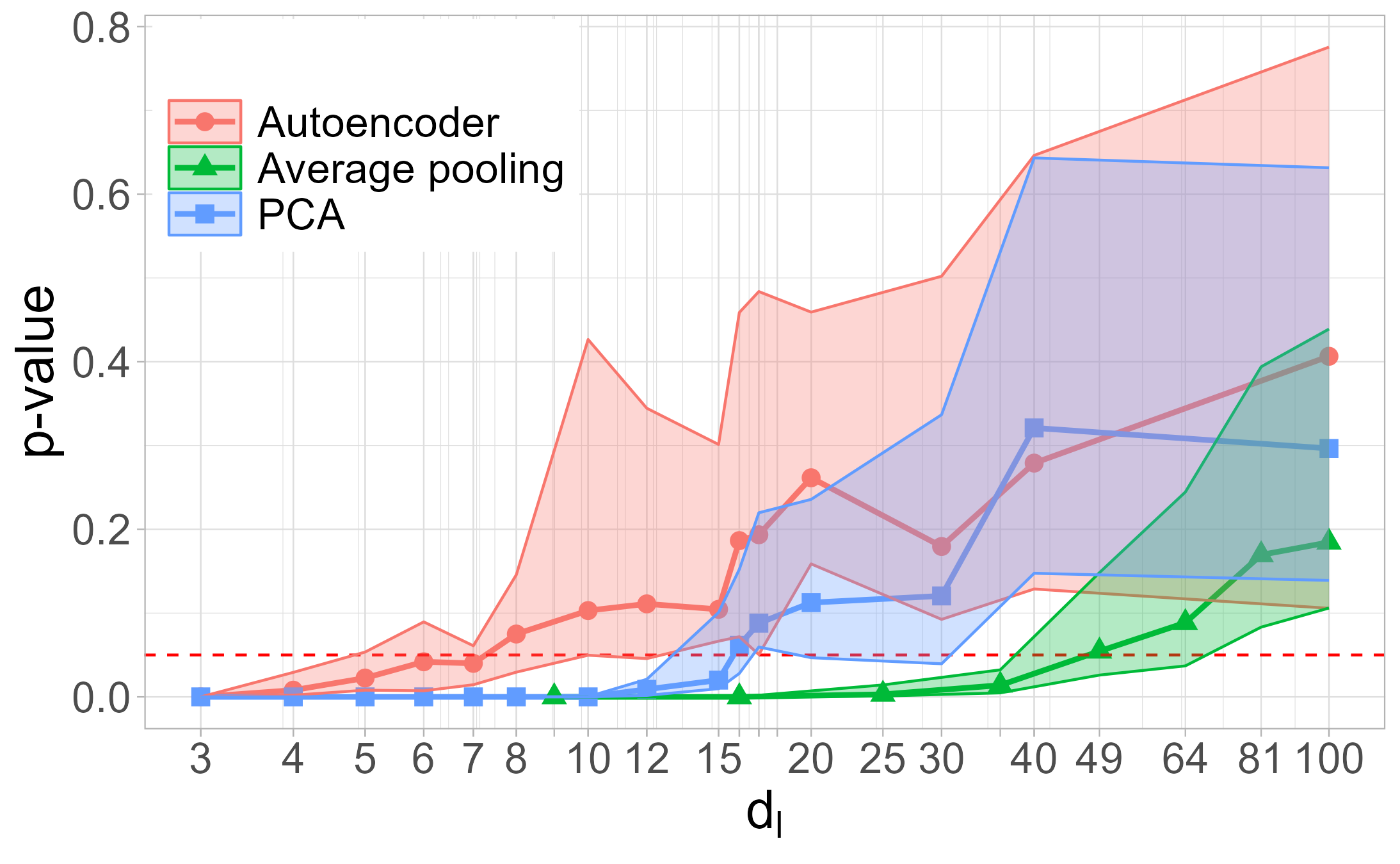}
		\caption{ \footnotesize The median p-values and its $25\%$ and $75\%$ quantiles for AE, average pooling, and PCA. }
		\label{fig_sec4_2_2}
	\end{figure}

The median of the 40 p-values are plotted in Figure \ref{fig_sec4_2_2} as a function of $d_l$. For AE, the median and  25\% quantile of the p-values exceeds 5\% level when $d_l\geq 7$ and $d_l\geq 10$ respectively. This suggests the dimension of feature vector is about $10$ to $15$, which is consistent with many empirical findings for MNIST data \citep{fournier2019empirical}. For PCA, the 75\% quantile of the p-values remains close to zero when $d_X\leq10$ and the median p-value rises above 5\% level when $d_l>15$. For average pooling, the 75\% quantile of the p-values remains close to zero even if the pooled image has $5\times 5$ pixels (implemented by torch.nn.AdaptiveAvgPool2d with kernel\_size${=}8$, stride${=}5$, padding${=}0$). As expected, this result demonstrates that AE is much more efficient than PCA and average pooling in feature extraction for MNIST dataset.
	  \normalsize

	\section{Conclusion}\label{sec6conclu}

	In this paper, we propose a novel conditional independence (CI) test that overcomes the challenges faced by existing tests when the dimensions of the data are large. By utilizing reproducing kernel Hilbert space (RKHS) embedding, sample splitting, cross-fitting as well as generative neural networks (GNNs), the test we developed is fully non-parametric, alleviates the curse of dimensionality, and is doubly robust against approximation errors of GNN generators. 
 Extensive simulations demonstrate the accurate size and satisfactory power performance of the proposed method in comparison with existing ones. The data examples further illustrate the versatility of our proposed method in real-world applications.
    
   To conclude, we discuss several potential extensions. In regression modeling and sufficient dimension reduction, it is often more useful to test the weaker null hypothesis of conditional mean or quantile independence to determine whether an additional predictor contributes to predicting the mean or quantile of the response variable given a set of covariates. Generalizing the proposed test to address these problems would be beneficial. Additionally, constructing a CI test with triple or multiple robustness properties for testing mutual conditional independence among multiple random variables may be desirable. Moreover, it would be interesting to apply the current method to several downstream statistical applications. For example, in time series model specification testing, where the Markov property described through conditional independence plays an important role \citep{zhou2023testing}, and in nonlinear dimension reduction problems, where our test procedure could be promising in identifying the optimal dimension reduction mapping by appropriately inverting the test statistic into an estimator in a computationally efficient manner; see \cite{huang2024reduc} for some recent work. We leave these topics for further research.

	\bibliographystyle{chicago}
	
	\bibliography{Bib_condi_indep}

\end{document}